\documentclass[aps,prx, superscriptaddress,twocolumn,longbibliography]{revtex4-2}

\usepackage{graphicx}
\usepackage{physics}
\usepackage{float}
\usepackage{hyperref,xcolor,comment}
\usepackage[normalem]{ulem}
\usepackage{amsmath}
\usepackage{amssymb}
\usepackage{amsfonts}
\usepackage[T1]{fontenc}
\usepackage{lmodern}

\begin{document}
\title{Inter-Species Interactions in Dual, Fibrous Gel Enable Control\\ of Gel 
Structure and Rheology}
\author{Mauro L Mugnai}
\email{mm4994@georgetown.edu}
\affiliation{Institute for Soft Matter Synthesis and Metrology, Georgetown University, 3700 O St NW, Washington, DC, 20057, USA}
\author{Rose Tchuenkam Batoum}
\affiliation{Institute for Soft Matter Synthesis and Metrology, Georgetown University, 3700 O St NW, Washington, DC, 20057, USA}
\affiliation{Department of Physics, Georgetown University, 3700 O St NW, Washington, DC, 20057, USA}
\author{Emanuela Del Gado}
\email{ed610@georgetown.edu}
\affiliation{Institute for Soft Matter Synthesis and Metrology, Georgetown University, 3700 O St NW, Washington, DC, 20057, USA}
\affiliation{Department of Physics, Georgetown University, 3700 O St NW, Washington, DC, 20057, USA}


\begin{abstract}
    Natural and synthetic multi-component gels display emergent properties, which implies that they are more than just the sum of their components.
    This warrants the investigation of the role played by inter-species interactions in shaping gel architecture and rheology.
    Here, using computer simulations, we investigate the effect of changing the strength of the interaction between two species forming a fibrous double network.
    Simply changing the strength of inter-species lateral association, we generate two types of gels: one in which the two components demix, and another one in which the two species wrap around each other.
    We show that demixed gels have structure and rheology that are largely unaffected by the strength of attraction between the components.
    In contrast, architecture and material properties of intertwined gels strongly depend on inter-species ``stickiness'' and volume exclusion.
    These results can be used as the basis of a design principle for double networks which are made to emphasize either stability to perturbations or responsiveness to stimuli.
    Similar ideas could be used to interpret naturally occurring multi-component gels.
\end{abstract}
\maketitle
Multi-component gels are prominent in nature and are becoming a frontier in materials design.
The space between and within human cells is permeated by two multi-component networks, respectively the extra-cellular matrix (ECM) and the cytoskeleton~\cite{AlbertsBook}.
The ECM ingredients are a tissue-dependent mixture of protein fibers and polysaccharide gels~\cite{Padhi2020AnnBioEng,Frantz2010JCS}.
Cells respond to ECM material properties such as stiffness and stress relaxation timescales, while ECM stiffness also changes cell differentiation, and its heterogeneous microenvironments contribute to morphogenesis~\cite{Discher2005Science, Chaudhuri2020Nature,Rozario2010DevBio}.
For instance, mixtures of collagen and polysaccharides (agarose or cross-linked hyaluronan) display material properties that emerge from a sophisticated interplay between multiple components~\cite{Ulrich2010Biomaterials,Burla2019NatPhys}.
The cytoskeleton, in turn, is also made of multiple fibrillar components: the filamentous actin (F-actin), microtubules (MTs), intermediate filaments (IFs) and septins~\cite{Pollard2018CSHPB,Mostowy2012NRMCB}.
{\it In-vitro} composites made of F-actin and MTs or F-actin and IFs display strain stiffening or tunable softening/strengthening which are absent in the single-component gels~\cite{Lin2011SoftMatter,Jensen2014BioArch}.

Under appropriate preparation protocols, synthetic gels formed of two intertwined polymer components (interpenetrating polymer networks, or IPNs) display remarkable materials properties~\cite{Gong2003AdvancedMat}. 
The combination of a stiff and brittle network with a soft and ductile counterpart underlies the striking increase in toughness of the combined networks, which far exceeds the summation of the properties of the two individual components~\cite{Gong2010SoftMat}. 
Similarly, combinations of elastomers with differences in their capacity to dissipate energy~\cite{Ducrot2014Science} synergistically give rise to materials of extreme toughness, with mechanical characteristics that do not exist in either of the two components separately~\cite{li2023SciAdv,yu2020AdvMater,li2024dnatrevmat}. The organization of the components in the composite network architecture, whether sequential or random, and not only their chemical composition, seems to be a determinant factor for these emergent mechanical properties~\cite{li2020PNAS, webber2007Macro}. 

In sum, existing natural or synthetic multi-network gels provide a concept for material design where the composite network is {\it more} than the sum of its components, and which points to role of {\it inter-}network interactions. Particularly interesting are  the implications of these ideas for self-assembling multi-component colloidal materials that can exploit advances in synthesis and self-assembly of functional nanoparticles \cite{glotzer2007anisotropy,Grzelczak2010directed,Chen2011triblock,Aida2012functional,Kotov2014self,travesset2015binary}. Interpenetrating colloidal gels can be obtained via arrested demixing of the different colloidal components~\cite{Varrato2012PNAS,Immink2019reversible} and the level of control already achieved in the design of nanoparticles surfaces allows for self-assembly of gel networks whose structural units can be as complex as fibrils, ramified objects, or chiral elements, rapidly expanding the frontiers for new material design \cite{Rijns2024,Kim2024direct,Mao2024MRS}. It is clear, however, that moving forward with novel nanoparticle network concepts, as well as understanding fully the functioning of already existing multi-network systems, require control of the {\it inter-}network interactions. 

Here, we hypothesize that {\it inter-}species interactions are crucial, on their own, to determine architecture and rheology for multi-component gels made of natural or synthetic networks, and directly test this hypothesis in a computational model for self-assembling multi-network colloidal gels. We focus on a binary mixture (i.e., two colloidal species), and, thanks to 3D computer simulations, systematically change inter-species interaction parameters to explore the outcomes from the micro- to the macro-scale. The goal is to link the inter-network interactions, and the resulting double-network spatial organization, to the emergent mechanical properties. For the sake of simplicity and clarity, we consider symmetric mixtures made of equal amount of two identical components that are described by the same intra-species interactions. These choices help us in fundamentally singling out the role of inter-species interactions. 

\begin{figure}
\centering
\includegraphics[trim=0cm 18cm 0cm 0cm,angle=0,width=0.5\textwidth]{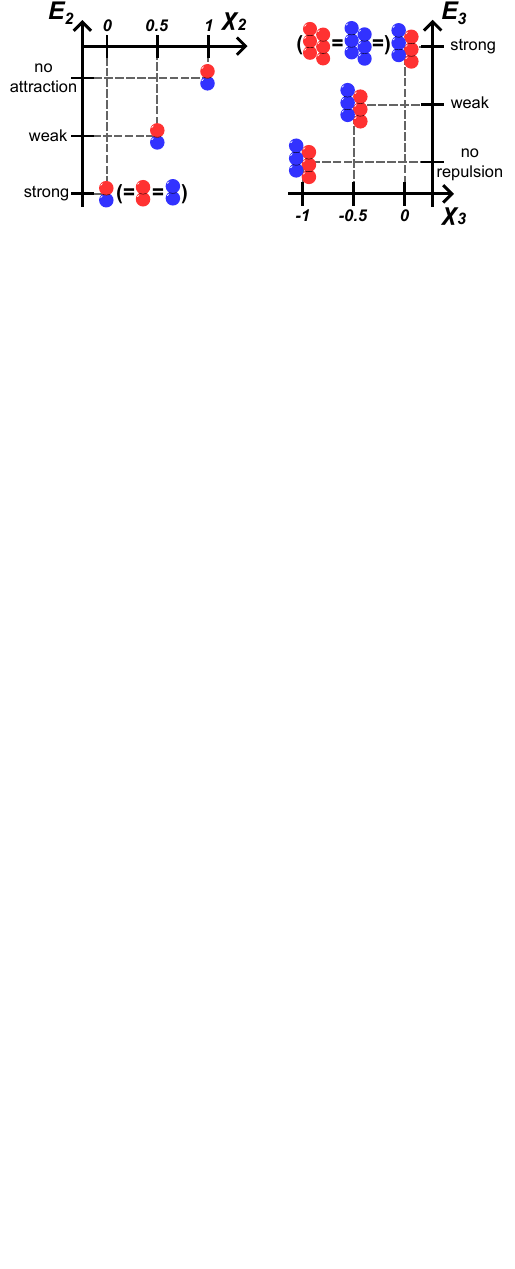} 
\caption{Pictorial illustration of how the parameters $\chi_2$ (left) and $\chi_3$ (right) change two-body ($E_2$) and three-body ($E_3$) inter-species interaction strength.
Particles of the two species are colored in red and blue.
}
\label{Fig:Model}
\end{figure}

We discover that the structures resulting from varying the inter-species interactions can be categorized into two groups: (i) composites undergoing {\it demixing}, in which the two species occupy separated compartments resembling a three-dimensional checkerboard, and (ii) {\it intertwined} networks, featuring branches of the two networks that are nearly colocalized and in some cases wrapped around each other. 
Monitoring a variety of structural and rheological parameters, we propose the following design principle for double networks: robustness and susceptibility to perturbations of the inter-species interactions can be encoded in the gel by fabricating demixing and intertwined networks, respectively.
Experiments can test our findings using the combination of morphological analysis, accessible via confocal microscopy, and rheology ~\cite{Rijns2024}. 
In addition, our results can aid the design of new double-gel-based materials applied to expanding fields such as tissue engineering and additive manufacturing, where the composite help set properties such as porosity and stiffness (crucial for cell survival)~\cite{fung2013biomechanics, moutos2007nautureM,lee2001ChemRev} and printability~\cite{Wallin2020NatComm,Romberg2023AdditMan}.

\begin{figure*}
\centering
\includegraphics[trim=0.0cm 16cm 0cm 0.0cm,angle=0,width=\textwidth]{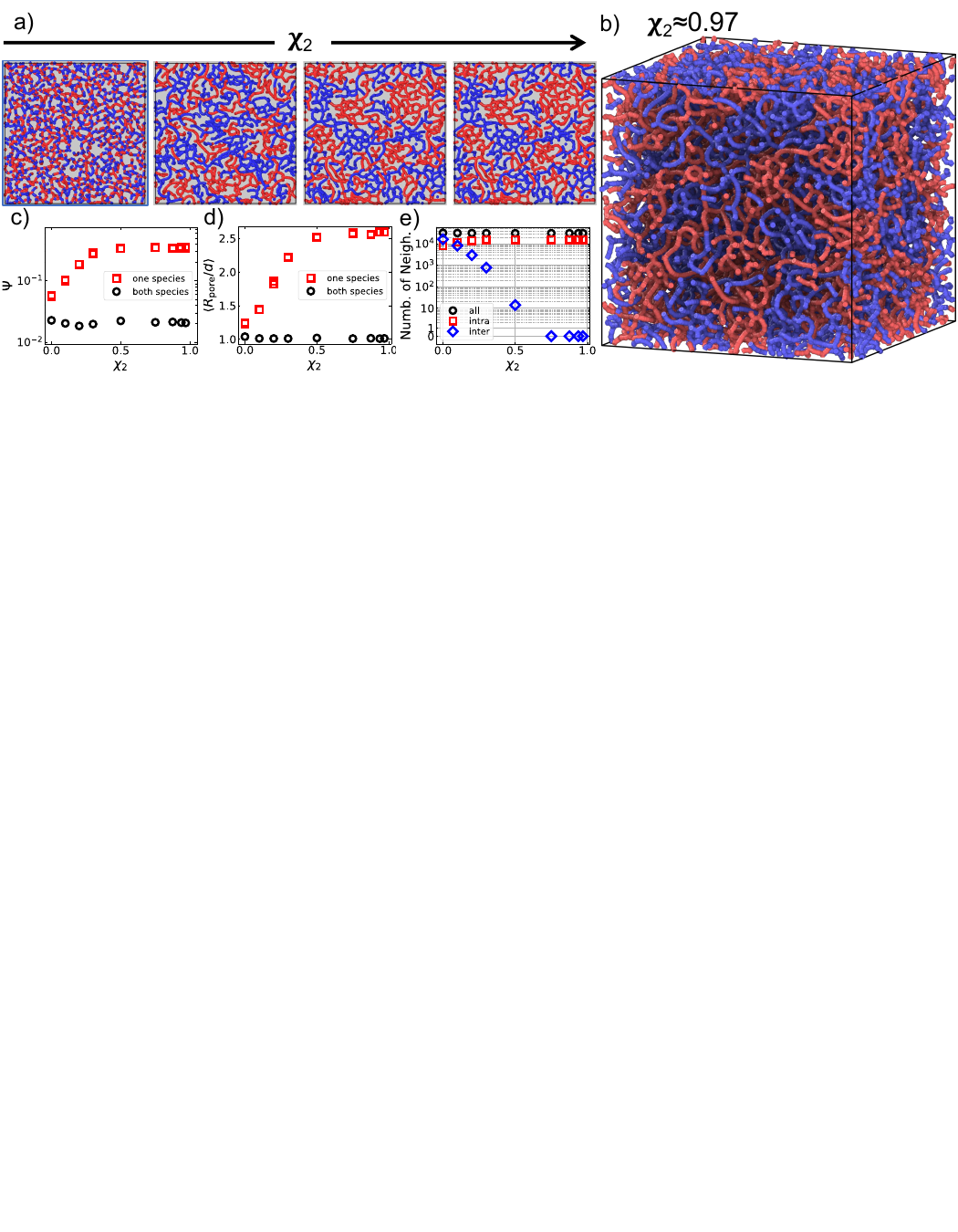} 
\caption{Demixing networks, $\chi_3=0$.
(a) Section of the three-dimensional gel of thickness $5d$.
Particles are shown as spheres connected by bonds if they are within $1.1d$ from each other, that is near the minimum of the pairwise potential.
From left to right, the four figures refer to $\chi_2=0$, $0.5$, $0.75$, and $0.875$.
(b) Full three-dimensional picture of the gel obtained with $\chi_2=31/32\approx0.97$.
(c) Demixing order parameter as a function of $\chi_2$.
Black circles and red squares show the results when all particles or only one species at a time was considered, respectively.
There are two nearly overlapped red squares per $\chi_2$, one per species.
(d) Same as (c), but showing the average pore size as a function of $\chi_2$.
(e) Number of contacts (distance $\le 1.5d$, which includes essentially the entire attractive well of the pairwise potential) between any two colloidal units (black circles), between colloidal units of the same species (red squares), and colloidal units of different species (blue diamonds).
}
\label{Fig:DemMorph}
\end{figure*}

\section*{Results}
In our model systems, each of the two components occupies the same volume fraction $\phi \approx 0.05$. The intra-species interactions are identical, and are given by a two-body radially symmetric term that enforces volume exclusion via a repulsion term that dominates at short distances, and a short-ranged attractive term that drives self-assembly. 
While the effect of this kind of interaction on bicontinuous assembly of multi-component gels has already been explored in previous computational studies~\cite{Varrato2012PNAS,FerreiroCordova2020SoftMatter}, we include in our model a second term that allows us instead to account for more complex interactions, such as those controlled by functional surface groups and surface heterogeneities in nanoparticle systems \cite{Bantawa2021JPhysCondMat}. This strategy allows us to focus on self-assembled gels with fibrous structures, as many gels, especially in the biological context, are made of spanning and fibrillar structures.
This second term in the energy function is modeled as a solely repulsive three-body term which favors the formation of filamentous structures that get connected into a spanning network even at low particle volume fractions \cite{Colombo2013PRL,Colombo2014SoftMatter,Bantawa2021JPhysCondMat}. This approach has been shown to lead to network structures that well recapitulate behaviors typical of a broad range of single-network colloidal gels, or composites made of a spanning network and small aggregates, in terms of their structural, dynamical, mechanical, and rheological characteristics~\cite{Colombo2014SoftMatter,Colombo2014JofRheology,Bouzid2017NatComm,Bouzid2018Langmuir,Bantawa2023hidden,Vereroudakis2020ACSCS,Donley2022JofRheol,Dellatolas2023local}. 
Here, we use it to explore instead the self-assembly of two identical network-forming colloidal species into a range of different dual morphologies, depending on their inter-species interactions. 
 
The strength of the inter-species interaction is given in terms of two parameters, $\chi_2$ and $\chi_3$ for two-body and three-body interactions respectively (see Eq.~\ref{Eq:chi2} and Eq.~\ref{Eq:chi3}), whose meaning is pictorially described in Fig.~\ref{Fig:Model}.
The parameter $\chi_2$ controls the strength of inter-species two-body attraction.
Drawing a connection with mean-field theories of phase separation (i.e. Bragg-Williams or Flory-Huggins~\cite{HillBook}), $\chi_2$ corresponds to half of the relative change in energy of the system when a pair of intra-species bonds is broken and replaced by two inter-species bonds. If $\chi_2=0$, there is no difference between inter- or intra-species pairwise attraction (see Fig.~\ref{Fig:Model}).
A positive value of $\chi_2$ means that inter-species bonds are less stable than intra-species ones, and hence creating them costs some energy.
The $\chi_2$ parameter is normalized so that if there is no inter-species interaction at all then $\chi_2=1$.
In other words, $1-\chi_2$ is a measure of inter-species stickiness, which is absent for $\chi_2=1$ and maximum (i.e. equivalent to intra-species interaction strength) if $\chi_2=0$.
It should be noted that increasing $\chi_2$ also slightly reduces volume exclusion (see Fig.~S1).
The parameter
 $\chi_3$, instead, which is again obtained using a three-body repulsive term (see Methods), quantifies the tendency of fibrils of different species to laterally associate and bundle (see Fig.~\ref{Fig:Model}).
When $\chi_3=0$, fibrils of different species cannot bundle, whereas a decrease in $\chi_3$ indicates a decrease in the repulsion that prevents lateral association of fibrils of different species, and for $\chi_3=-1$ there is no penalty for bundling.

In the following, we focus on: (i) networks generated for $0\le \chi_2<1$ at $\chi_3=0$, in which the two-body inter-species attraction is varied while the lateral association of the fibers of different species is disfavored, and (ii) gels made with $\chi_3=-1$ and $0\le\chi_2<1$, in which instead the fibers of distinct species can lay side-by-side or wrap around each other owing to an attractive energy regulated by $\chi_2$.
As we show, the first energy function (i) gives rise to double networks in which the two components occupy distinct compartments, and therefore we term these ``demixed networks''.
The second energy function (ii) produces gels in which there are instances of fibers of the two components wrapped around each other, and thus we define them ``intertwined networks''.
Besides presenting different morphologies, demixed and intertwined gels display a strikingly different dependence on $\chi_2$ of their response to mechanical perturbation and to changes in the parameters of inter-species interactions after the gel is set. 
\begin{figure*}
\centering
\includegraphics[trim=0cm 16cm 0cm 0cm,angle=0,width=\textwidth]{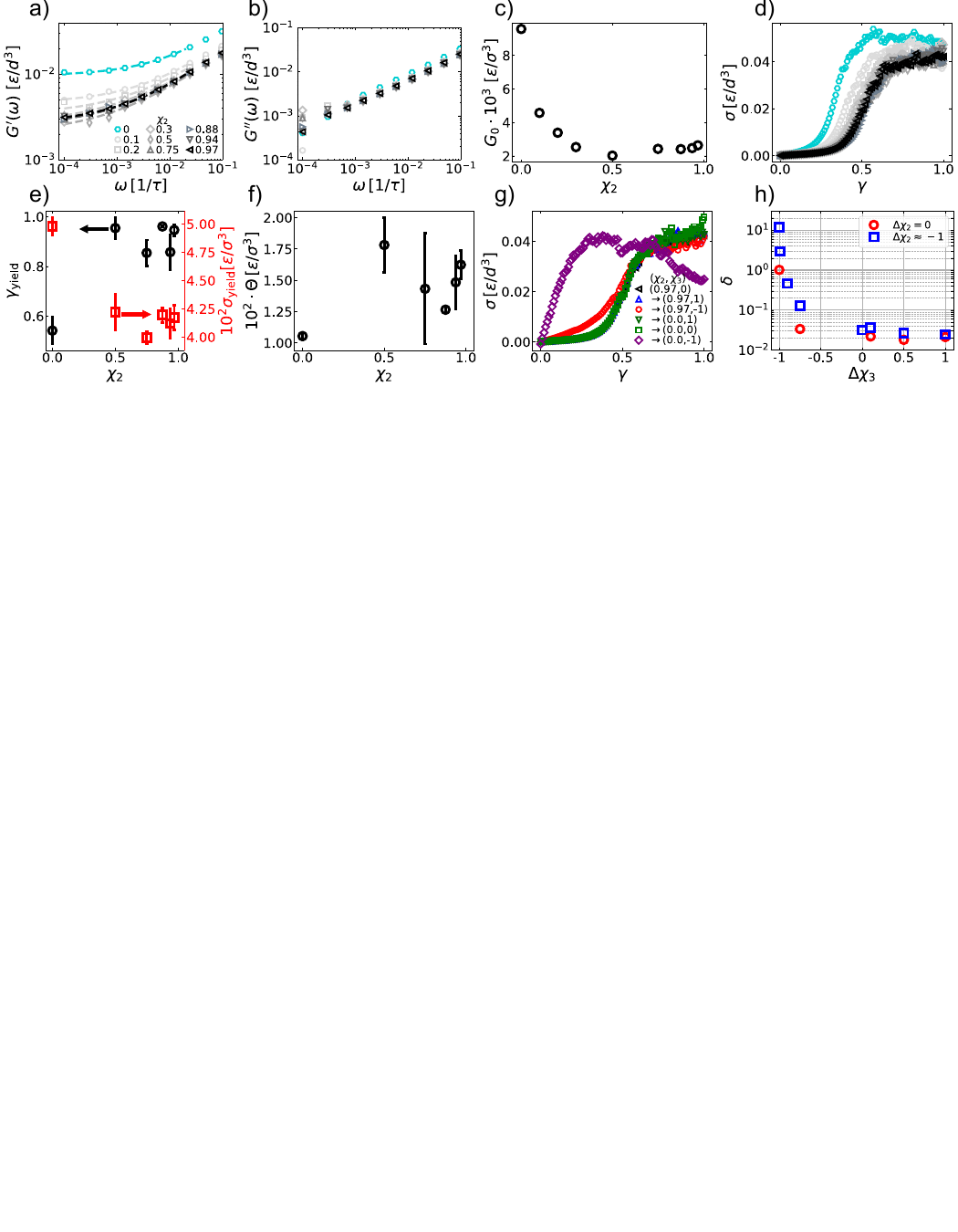}
\caption{Rheology of demixing gels ($\chi_3=0$).
(a-b) Linear oscillatory rheology.
Panel (a) displays the elastic moduli, the loss moduli are shown in panel (b).
In both panels, data are shown as empty symbols of different shapes and colors representing different $\chi_2$ values (see legend where the values are rounded to the second significant digits, the exact values are $0$, $0.1$, $0.2$, $0.3$, $0.5$, $0.75$, $0.875$, $0.9375$, and $0.96875$).
For clarity, we show only a subset of data points.
The dashed lines show the fit of the low-frequency domain of the spectrum to $G_0 + K\omega^\alpha$, where $G_0$, $K$, and $\alpha$ are fitting parameters.
(c) Elastic modulus obtained from the fit in panel (a) as a function of $\chi_2$.
(d) Non-linear rheology, start-up shear.
The colors and symbols indicate the value of $\chi_2$ and are reported in the legend in panel (a).
(e) Yield strain (black circles, left axis) and stress (red squares, right axes) as a function of $\chi_2$, obtained from the maximum of the load curves.
(f) Toughness of the gel, $\Theta$, defined as the integral of the stress as a function of strain up to the yielding point.
(g) Load curve after perturbation of the inter-species interaction parameters.
The reference is always $\chi_2\approx0.97$ and $\chi_3=0$, the new parameters are shown in the legend after an arrow.
(h) Change in the load curve after the perturbation.
The value of $\delta$ indicates the change in the load curve between before and after the perturbation, relative to the unperturbed load curve. 
Changes in $\chi_3$ are shown in the x-axis, the legend indicates changes in $\chi_2$.
}
\label{Fig:DemRheo}
\end{figure*}

\subsection*{Demixed Networks: $\chi_3 = 0$}
\paragraph{Structure of the Network:} 
Figures~\ref{Fig:DemMorph}a-b display structures produced in our simulations using various values of $\chi_2$.
In the nearly two-dimensional slices of the three-dimensional (3D) system (Fig.~\ref{Fig:DemMorph}a) as well as in the 3D snapshot (Fig.~\ref{Fig:DemMorph}b), one species is shown in red and the other one in blue.
In the first slice of Fig.~\ref{Fig:DemMorph}a both $\chi_2=0$ and $\chi_3=0$, therefore the distinction between the two species is only formal, all the particles are identical, and thus the gel is a single-network of volume fraction $\phi\approx0.10$ with red and blue particles completely interspersed.
Instead, it is clear that increasing $\chi_2$ in Figs.~\ref{Fig:DemMorph}a leads to the partial compartmentalization of the two components.
The first question to ask is: are these double networks, in which each component has independently percolated, or is this a two-component single-network, where only the whole system is spanning?
For $\chi_2 < 0.2$, only the cluster made by both species is space-spanning, and thus we consider such systems as single networks; instead, for $\chi_2 \ge 0.2$ each species percolates on its own, giving rise to a double network.
Second, we wish to quantify the visual observation that the two networks are compartmentalized. 
In order to do so, we compute the demixing order parameter for a $N$-particle system, $\Psi$, which monitors the spatial fluctuations of particle number density from the average computed in $10^3$ cubic subdivisions of the simulation box (see Supporting Information).
For a homogeneous system $\Psi \approx 0$, whereas if only half of the cells are populated $\Psi \approx 1$.
Figure~\ref{Fig:DemMorph}c shows the demixing parameter as a function of $\chi_2$ computed either accounting for all colloidal units regardless of the species (black circles) and for one species at a time (red squares).
When all colloidal units are considered, $\Psi$ is small and nearly completely independent of $\chi_2$.
This suggests that globally the volume is spanned uniformly by demixed fibril-forming networks, regardless of inter-species ``stickiness'' ($1-\chi_2$).
Similarly, Fig.~\ref{Fig:DemMorph}d shows that the average pore size between all colloidal particles is nearly independent on $\chi_2$.
Focusing on one of the two species at a time, both the demixing order parameter (Fig.~\ref{Fig:DemMorph}c, red squares) and the average pore size (Fig.~\ref{Fig:DemMorph}d, red squares) increase with $\chi_2$, suggesting the increase in spatial segregation between the two species which saturates for $\chi_2 \ge 0.5$.
This suggests that the inter-species interactions become negligible as $\chi_2$ increases.
Indeed, while the total number of contacts (Fig.~\ref{Fig:DemMorph}e, black circles) and the intra-species contacts (red squares) are nearly independent on the strength of inter-species ``stickiness'' ($1-\chi_2$), the number of inter-species contacts at $\chi_2=0.5$ is three orders of magnitude lower than it is for $\chi_2=0$, and for $\chi_2 > 0.5$ the two components are completely segregated.
Overall, Fig.~\ref{Fig:DemMorph}e suggests that the saturation of demixing and average pore size correlate with the disappearance of inter-species contacts.

\paragraph{Linear Oscillatory Rheology:} Using the Optimally Windowed Chirp (OWCh) method~\cite{Geri2018PRX,Bouzid2018JofRheology}, we extracted the storage ($G^\prime$) and loss ($G^{\prime\prime}$) moduli for all of the networks (see Fig.~\ref{Fig:DemRheo}a,b, Fig.~S2a,b, and details in the SI).
While the loss moduli are nearly independent on the strength of two-body interactions, the storage moduli at low frequency suggest that the shear modulus [$G_0=G^\prime(\omega=0)$] depends on $\chi_2$.
To extract $G_0$, we fit the low-frequency region of the storage modulus to $G_0 + K \omega^\alpha$, where the exponent $\alpha$ is also a parameter of the fit (this equation is the low-frequency limit of the expression provided by Bouzid {\it et al}~\cite{Bouzid2018JofRheology}).
The shear modulus saturates for $\chi_2 \gtrapprox 0.3$ (Fig.~\ref{Fig:DemRheo}c), suggesting that weakening the inter-species interactions below a threshold has little to no impact on the linear rheology of the gel.

\begin{figure*}
\centering
\includegraphics[trim=0cm 16cm 0cm 0cm,angle=0,angle=0,width=\textwidth]{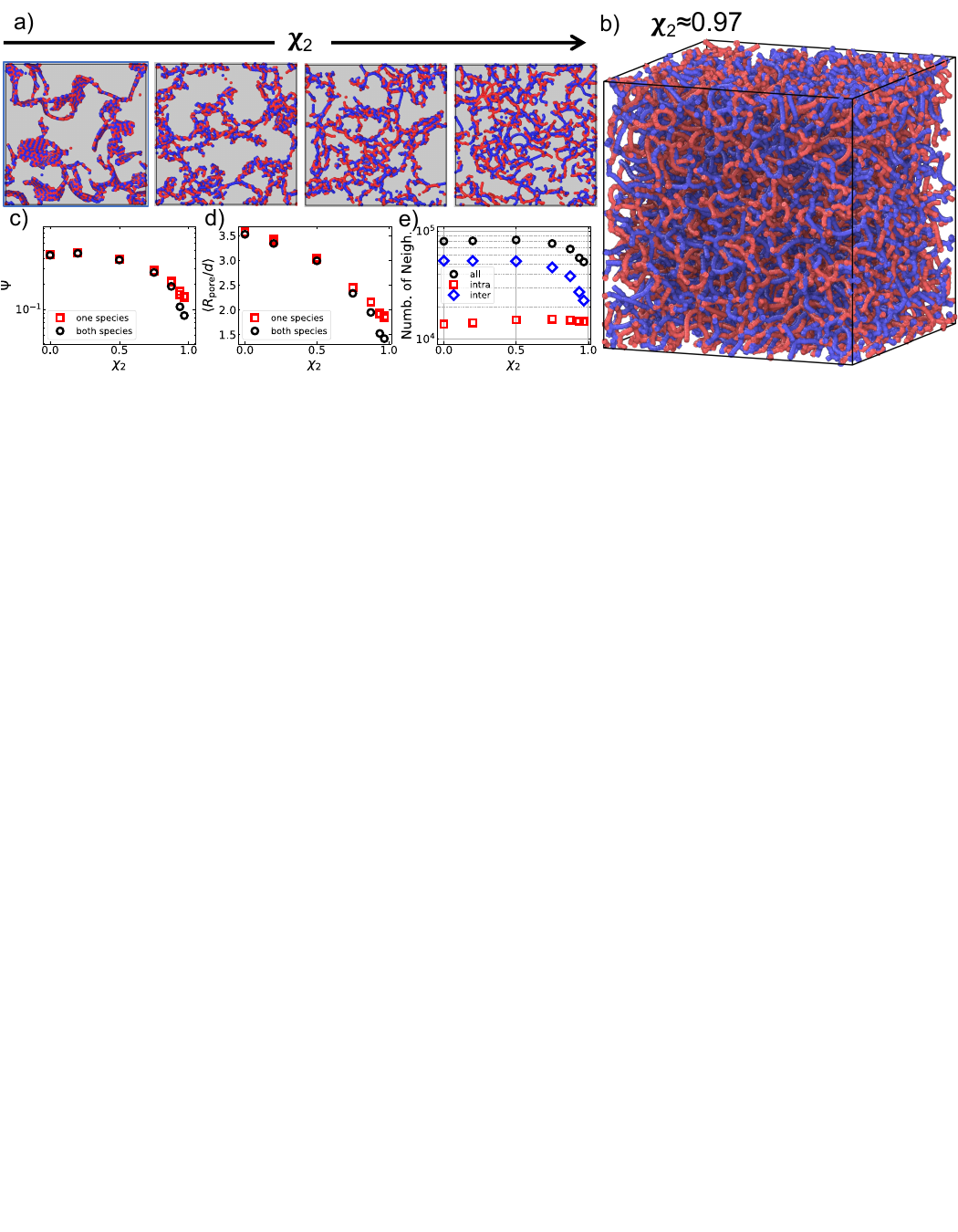} 
\caption{Structure of intertwined gels ($\chi_3=-1$).
(a) Sections of the gels of thickness $5d$.
Particles (spheres) and bonds (for pairs within $1.1d$ from each other, close to the minimum of the pairwise potential) are shown in blue or red to identify the two species.
Going from left to right (see arrow) the $\chi_2$ parameter for each slice is $0$, $0.5$, $0.75$, and $0.875$.
(b) Three-dimensional representation of the gel obtained using $\chi_2=31/32\approx0.97$.
(c) Demixing order parameter for all colloidal units (black circles) or for one components at the time (red squares).
(d) Average pore size in the network -- colors and symbols are the same as in panel (e).
(e) Number of contacts (distance $<1.5d$, after which the pairwise potential is effectively negligible) between all units (black circles), between units of the same species (red squares), and between units of different species (blue diamonds)
}
\label{Fig:MorphInter}
\end{figure*}

\paragraph{Start-Up Shear:}
Figure~\ref{Fig:DemRheo}d shows the calculation of the load curves for all the gels belonging to the demixing class ($\chi_3=0$).
The stress is calculated using a start-up shear protocol~\cite{Colombo2014JofRheology} (see SI for details).
The curves display a common shape and display robust strain stiffening with onset depending on the strength of the two-body interaction up to $\chi_2 \approx 0.5$, after which the load curves become independent on $\chi_2$ (Fig.~\ref{Fig:DemRheo}d).
Similarly, the strain and stress at which the network yields, $\gamma_\mathrm{yield}$ and $\sigma_\mathrm{yield}$, which we define from the maximum of the load curve (see SI for details), change and then saturate as $\chi_2$ grows (Fig.~\ref{Fig:DemRheo}e).
The toughness, $\Theta$, defined as the integral of the load curve up to the yield point ($\Theta = \int_0^{\gamma_\mathrm{yield}}d\gamma \sigma$, see SI for details), is larger than for a single network ($\chi_2=0$) but essentially constant for $0.5\le\chi_2<1$ (Fig.~\ref{Fig:DemRheo}f, black circles).

\paragraph{Post-Gelation Modifications:} 
Given that the load curve becomes independent of $\chi_2$, we hypothesize that the network should be robust to changes of the parameters even {\it after} gelation.
To test this conjecture, we perturb a self-assembled multi-component gel through the following procedure: (i) we select some initial parameters $\chi_2$ and $\chi_3$ and produce a gel as done so far; (ii) we change the strength of inter-species interaction to $\chi_2^\prime$ and $\chi_3^\prime$; (iii) we allow the network to adjust to these new parameters, and (iv) we monitor the load curve of the modified double gel.
As a measure of the changes in the load curve, we define $\delta$ as the relative error between the stress computed for the perturbed ($\sigma_\mathrm{pert}$) and un-perturbed ($\sigma_\mathrm{unpert}$) networks averaged over all the strains ($\delta = 1/N\sum_{i=1}^N |\sigma_\mathrm{pert}(\gamma_i)-\sigma_\mathrm{unpert}(\gamma_i)|/|\sigma_\mathrm{unpert}(\gamma_i)|$).
As initial parameters we set $\chi_2\approx0.97$ and $\chi_3=0$.
As it is clear from Fig.~\ref{Fig:DemRheo}g, there is no significant change in the load curve unless inter-species fibril bundling is allowed ($\chi_3=-1$, $\Delta \chi_3 = -1$) {\it and} the inter-species attraction is restored to the inter-species value, i.e. $\chi_2=0$ and $\Delta \chi_2 \approx -0.97$. 
The results are summarized and quantified in Fig.~\ref{Fig:DemRheo}h, where it is clear that demixed networks display robustness to the strength of inter-species interaction. 

In sum, gels formed with $\chi_3 = 0$ give rise to double networks only if attractive inter-species interactions are sufficiently weak ($\chi_2 \ge 0.2$).
For these double networks (i) morphology, characterized by the demixing order parameter and the average pore size, and (ii) rheology, defined by the elastic modulus, yielding and toughness, are only weakly if at all dependent on the strength of inter-species interaction.
Finally, (iii) demixed networks display robustness to post-gelation perturbation of the interactions.

\subsection*{Intertwined Networks: $\chi_3 = -1$}
\paragraph{Structure of the Network:} 
The absence of three-body inter-species repulsion ($\chi_3=-1$) enables bundling between colloidal fibers of different species (see Fig.~\ref{Fig:Model}).
This is indeed what we observe in our simulations: the structures in Figs.~\ref{Fig:MorphInter}a,b show that red and blue units are well interspersed, in striking difference with the structures obtained when $\chi_3 = 0$.
Interestingly, for all the $\chi_2$ values considered here the gels are double networks, i.e. each species forms a percolating, spanning network. 
 For $\chi_2=0$ (first slice from the left in Fig.~\ref{Fig:MorphInter}a), homogeneous filaments of different species form semi-crystalline, nearly two-dimensional sheets, leaving large holes between the planes.
For $\chi_2=0.5$ (second slice from the left in Fig.~\ref{Fig:MorphInter}a), the planar arrangement is replaced by intertwined, twisted branches made of units of both components; visually, the architecture leaves plenty of holes between the branches. We note that increasing $\chi_{2}$ locally induces a small decrease of the inter-species excluded volume (Fig. S1), which may affect the local packing but does not explain the general trends. As $\chi_2$ increases, pairs of filaments emerge that are less tightly laterally associated, but are still intertwined and at times are wrapped around each other, in an arrangement that fills the space more uniformly (Fig.~\ref{Fig:MorphInter}a,b).
These observations are substantiated by the demixing order parameter (Fig.~\ref{Fig:MorphInter}c) and the average pore size (Fig.~\ref{Fig:MorphInter}d) as a function of $\chi_2$.
Indeed, if the demixing order parameter and the average pore size decrease as $\chi_2$ is increased, then the strength of inter-species interaction controls the porosity of the network, which is nearly the same if one species (red squares) or both species (black circles) are considered, again reinforcing the idea that the two networks are closely intertwined.
Furthermore, neither the demixing order parameter nor the average pore size saturate with $\chi_2$.
This suggests that even though the inter-species interaction is very weak, the packing and therefore contacts between the two species play an important role in regulating network architecture.
To confirm this, we monitor the contacts between colloidal units (Fig.~\ref{Fig:MorphInter}e).
The reduction of total contacts as $\chi_2$ increases (black dots in Fig.~\ref{Fig:MorphInter}e) is consistent with the increasingly more homogeneous occupation of space suggested by the demixing order parameter and average pore size.
As the number of intra-species contacts remains nearly identical at all $\chi_2$ values (red squares in Fig.~\ref{Fig:MorphInter}e), these changes are driven by the reduction of inter-species contacts (blue diamonds in Fig.~\ref{Fig:MorphInter}e) as the strength of two-body attraction between the two components is decreased.
Notably, not only the number of contacts between the two species is still large even at $\chi_2 \approx 0.97$, but more importantly it has not yet saturated, which confirms that the network architecture is still sensitive to the attraction strength between the two species even though it is now only about $3\%$ of the intra-species attraction strength.

\begin{figure*}
\centering
\includegraphics[trim=0cm 16cm 0cm 0cm,angle=0,width=\textwidth]{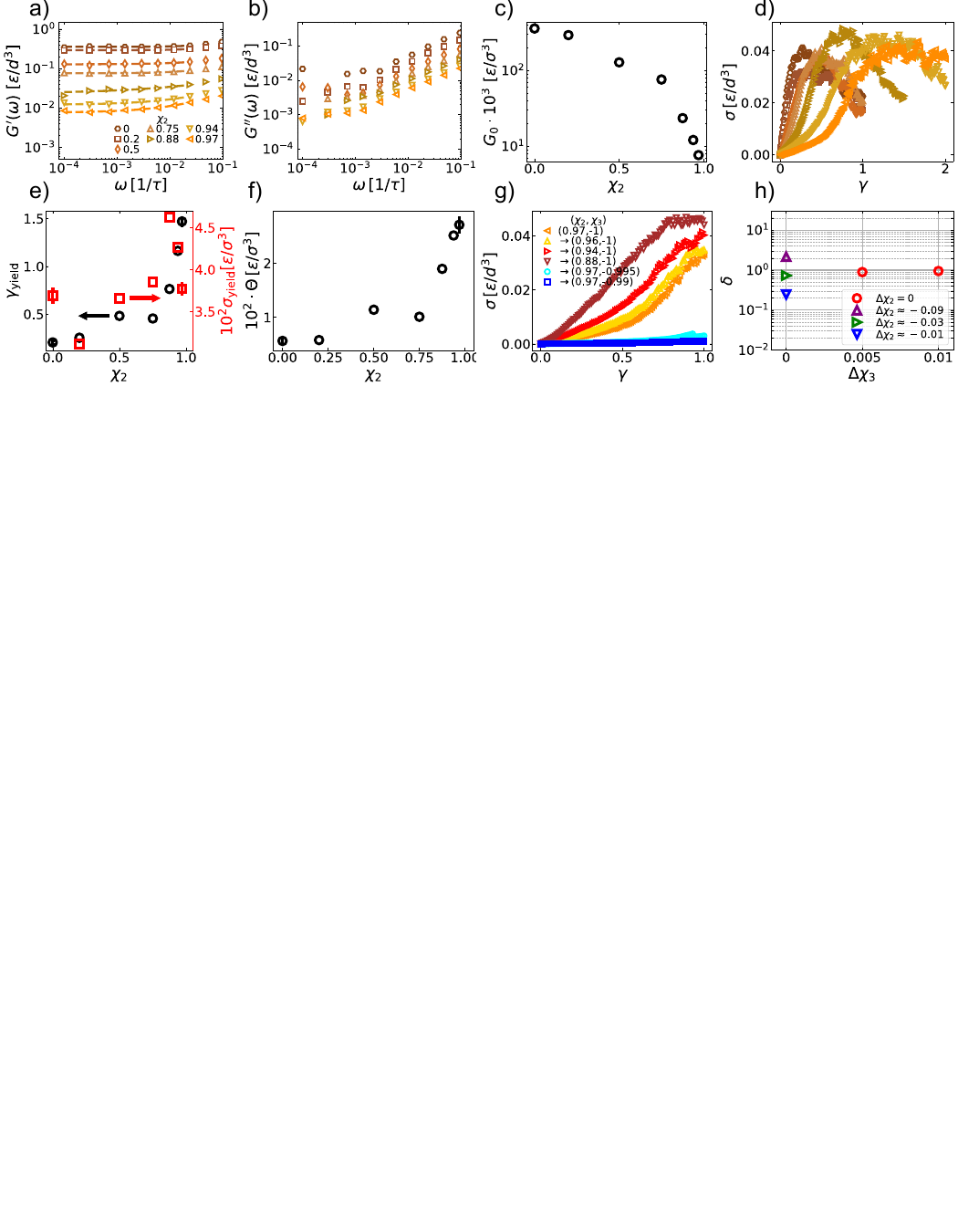}
\caption{Rheology of the intertwined gels ($\chi_3=-1$).
(a-b) Linear rheology results for the elastic moduli showing the elastic (a) and loss moduli (b) as a function of frequency ($\omega$).
The color code and the symbols in the legend indicate the value of $\chi_2$.
In the legend the values are rounded to the second significant digit, the exact values are $1-1/2^n$, where $n=0,1,2,3,4,5$.
Dashed lines in (a) show the fit to the low-frequency domain to $G_0+K\omega^\alpha$.
(c) Shear modulus (red squares, right axis) obtained by fitting the data in panel.
(d) Non-linear, start-up rheology.
Stress ($\sigma$) as a function of strain ($\gamma$) for all gels considered. 
The symbols and colors represent different values of $\chi_2$ [see legend in panel (a)].
(e) Dependence on $\chi_2$ of the strain (black circles, left axis) and stress (red squares, right axes) at the yielding point.
(f) Toughness (integral of the loading curve to the yielding point) as a function of $\chi_2$.
(g) Load curves for the gels modified after gelation of the network prepared with $\chi_2\approx0.97$ and $\chi_3=-1$ (orange triangles). In the legend, the arrow indicates that the energy function has been changed after gelation to the pair of values pointed to by the arrow. 
(h) Relative error ($\delta$) between the stress after perturbation of the energy function and the stress for unperturbed parameters during shearing.
The results are averaged over all the shear values (up to $\gamma=1.0$) during the start-up shear simulations.
The values of the $\chi_2$ and $\chi_3$ perturbations ($\Delta\chi_2$ and $\Delta\chi_3$) are shown in the legend and in the x-axis, respectively.}
\label{Fig:RheoInter}
\end{figure*}

\paragraph{Linear Oscillatory Rheology} 
The response to linear oscillatory tests performed using the OWCh protocol shows that as the strength of two-body inter-species interaction is reduced, both the storage (Fig.~\ref{Fig:RheoInter}a) and loss (Fig.~\ref{Fig:RheoInter}b) moduli of the gel decrease, indicating that for these systems $\chi_2$ tunes both the elasticity and the energy dissipation.
As shown in Fig.~\ref{Fig:RheoInter}c, the shear modulus extracted from the fit to the elastic modulus is sensitive to the inter-species attraction at least up to $\chi_2 \approx 0.97$.

\paragraph{Start-up Shear:} 
As $\chi_2$ increases, the stress vs strain curves display qualitatively different behaviors (Fig.~\ref{Fig:RheoInter}d).
When the inter-species interactions are strong, the curves show no appreciable strain-stiffening before the yielding point is reached.
As $\chi_2$ increases, so does the onset of strain stiffening.
Similarly, while the yield stress remains nearly constant, the yield strain increases with $\chi_2$ without showing signs of saturation (Fig.~\ref{Fig:RheoInter}e).
Accordingly, the double networks becomes tougher as $\chi_2$ increases, again showing no sign of saturation up to $\chi_2 \approx 0.97$ (Fig.~\ref{Fig:RheoInter}f).
The gel obtained with $\chi_2 = 0.97$ is nearly three times tougher than the single gel and about twice as tough as the demixed gels (Fig.~\ref{Fig:DemRheo}f).

\paragraph{Post-Gelation Modifications:}
Given the sensitivity of the networks to the inter-species interaction set {\it before} gelation, we hypothesizes that changing the parameters {\it after} gelation but before rheological tests would have a strong impact on the load curve.
Indeed, a small increase of interspecies attraction (reduction of $\chi_2$, so $\Delta\chi_2<0$, see red circles in Fig.~\ref{Fig:RheoInter}g) or slight reduction of inter-species bundling propensity (higher $\chi_3$, $\Delta\chi_3 > 0$, blue squares in Fig.~\ref{Fig:RheoInter}f) engenders a noticeable change of the stress-strain curve.
As summarized in Fig.~\ref{Fig:MorphInter}h, which shows the change between the load curves before and after perturbation relative to the unperturbed stress-strain curve, even small changes of $\chi_2$ or $\chi_3$ give rise to sizable changes in the load curves, i.e. $\delta \approx 1$.

In sum, gels formed with $\chi_3 = -1$ always produce double networks, regardless of $\chi_2$.
The inter-species two-body interaction regulates (i) the spatial organization of the two components, (ii) network porosity, (iii) elastic and loss moduli, (iv) the shape of the load curve, and (v) toughness even for large values of $\chi_2$.
Moreover, (vi) the gels are sensitive to post-gelation modifications of the inter-species interaction strength. 

\section*{Discussion}
The results obtained demonstrate that inhibiting inter-species bundling ($\chi_3=0$) gives rise to networks that segregate into checkerboard-patterned domains as $\chi_2$ increases (Fig.~\ref{Fig:DemMorph}a-b). Structure and rheology of these demixed gels are insensitive to changes in $\chi_2$ as long as $\chi_2 \ge 0.5$ (Fig.~\ref{Fig:DemMorph}c-e and Fig.~\ref{Fig:DemRheo}a-f).
Even more, the rheology of demixed networks displays robustness to post-gelation changes of the interaction parameters (Fig.~\ref{Fig:DemRheo}g-h).
In contrast, enabling inter-species bundling ($\chi_3=-1$) gives rise to gels in which fibers of the two species are intertwined (Fig.~\ref{Fig:MorphInter}a-b).
For these gels, $\chi_2$ tunes structure and rheology of the material without showing signs of saturating up to $\chi_2 \approx 0.97$ (Fig.~\ref{Fig:MorphInter}c-e and Fig.~\ref{Fig:RheoInter}a-f). The networks rheology is strongly affected by post-gelation changes in the parameters of the model (Fig.~\ref{Fig:RheoInter}g-h).
We propose the following rationalization for our observations. 
Demixed networks, being partially segregated, have fewer inter-species contacts as compared to co-localized, intertwined double gels. Therefore, it is reasonable that demixed networks should be more tolerant to changes in the inter-species interaction parameters.

In the model, intra-species interactions are kept fixed, while we explored the properties of the materials as a function of inter-species interactions.
This is an idealization that is only possible in simulations, which are thus the ideal tool to specifically disentangle the effect of inter-species interactions.
The implications are, for both natural and synthetic double network materials, that the details of inter-species interactions are a key ``knob'' to design or rationalize their properties. 

\section*{Conclusions}
Motivated by the emergence of new properties in multi-component gels, in this work we formulate the following hypothesis: inter-species interactions on their own can tune structure and rheology of composite gels.
We specifically tested this conjecture by designing a two-component fibril-forming colloidal gel in which we explored the properties of the materials as a function of inter-species interactions while keeping the intra-species interactions fixed.
We conclude that inter-species interactions on their own modulate the structural and rheological properties of the double gels. 

We suggest that this observation will be useful for the design of multi-component colloidal gels engineered to tune structural (e.g. porosity) and rheological (e.g. toughness) properties, or to be either robust or susceptible to change in preparation protocol and post-gelation actions that modify the inter-species interaction, for instance solvent changes or the introduction of cross-linkers.

The connection between architecture and materials properties of composite gels that emerges from our work could be experimentally tested using a combination of confocal microscopy, to access structural information, and rheology, in order to study material properties of the gels. Changing the properties of each individual species, exploring the effect of composition and of preparation protocol (sequential versus simultaneous gelation) on the synergy between the constituents of the composite are among the many possible avenues of future research.

We surmise that our findings will be helpful in various active fields of research that necessitate understanding of or benefit from multi-component gels, including ECM engineering~\cite{Lou2022NatRevChem}, bioprinting~\cite{Chimene2020AdvMat}, and the design of tough hydrogels~\cite{li2024dnatrevmat}.

\section*{Methods}
The model and gel preparation have been described before~\cite{Colombo2014JofRheology}.
Here, we briefly introduce the minimum information necessary to understand the $\chi_2$ and $\chi_3$ parameters for the double gel.
Further details on the model, calculation of demixing order parameter, porosity, linear and non-linear rheology are in the Supporting Information.
All simulations are performed in version of LAMMPS~\cite{Thompson2022CompPhysComm} suitably modified to include this energy function. 

\paragraph{Units}
Results are shown in internal units, with the mass ($m$), length ($d$) and energy ($\epsilon$) respectively corresponding to the colloidal mass, diameter, and intra-species bond energy.
Other units are derived from these three together with fundamental constants.
The energy function consists of two terms of finite range: (i) a spherically-symmetric two-body term featuring short range volume exclusion and a short attractive tail, and (ii) a purely repulsive three-body term controlling the angle between a triplet of neighboring colloidal particles.
The depth of the attractive potential is regulated by a parameter $A_{ij}$, a second parameter $B_{ij}$ tunes the strength of the repulsive three-body term, where $i$ and $j$ identify the species involved in the interaction.

\paragraph{Two Species} 
We keep $A_{ii}$ and $B_{ii}$ constant and equal for the two components.
We explore the dependence of system structure and rheology on two parameters: $\chi_2$ and $\chi_3$, which identify the changes in the strength of two-body and three-body inter-species interactions relative to the corresponding intra-species parameters.
We define $\chi_2$ as,
\begin{equation}
\chi_2 = 
\frac{A_{ii}-A_{ij}}{A_{ii}} = 1-\frac{A_{ij}}{A_{ii}}.
\label{Eq:chi2}
\end{equation}
and $\chi_3$ as,
\begin{equation}
\chi_3 = \frac{B_{ij}-B_{ii}}{B_{ii}} = \frac{B_{ij}}{B_{ii}}-1
\label{Eq:chi3}
\end{equation}
There is a change in the sign of $\chi_2$ and $\chi_3$, which reflects the idea that while two-body interactions include attraction, the three-body potential is always repulsive.

\paragraph{Gel Preparation}
In the sample preparation, the gelation of the two components is simultaneous and occurs in presence of thermal fluctuations. The kinetic energy is then drained in order to find a local inherent structure~\cite{Stillinger1995Science} of the gel, corresponding to a mechanically stable state. 
We study the changes in the structure of the networks and the stress generated in response to a shear deformation~\cite{LeesEdwards} in athermal conditions, to focus on the effect of the mechanical perturbation.
For all networks, we test whether we reached a stable state by monitoring the temperature of the system, which is below $10^{-9}\epsilon/k_B$ at the end of the preparation.
In a few cases, we repeated the preparation protocol $4$ times using much longer protocols, leading to temperatures consistently below $10^{-17}\epsilon/k_B$.
The load curve of these repeats is nearly indistinguishable from the original one (see Fig.~S3).
We have used these repeats to gather statistics and provide error bars for noisy observables (yielding and toughness).
For other observables (porosity) the results from the original structure and the average of the repeats are nearly the same, and the error bar are commensurate with or below the size of the symbols used in figures.

\paragraph{Modification of the Energy Function}
We tested the architecture robustness to changes in the energy function.
In order to do so, we started from a structure obtained using the above protocol, we modified the energy function and relax the system to find a local minimum.
We verified in all cases that again the temperature reached values around or below $10^{-9}\epsilon/k_B$.

\section*{Supporting Information Appendix (SI)}
Supporting Information (SI) contain a more detailed description of the Methods and a figure showing how the two-body potential is tuned by $\chi_2$.
We also report the linear rheology results with all data points up to $\omega = 0.1/\tau$ in order to show the fluctuations of the data, and the load curves with multiple repeats.\\

\section*{Acknowledgements}
MLM acknowledges the support of the ISMSM-NIST postdoctoral fellowship at Georgetown University. MLM thanks Dr. Minaspi Bantawa for providing LAMMPS scripts and a code used to prepare the double network. EDG acknowledges support from U.S. National Science Foundation (grant DMREF CBET-2118962) and the donors of the American Chemical Society Petroleum Research Fund (New Directions Grant 62108-ND6).

\bibliography{bibliography}

\begin{thebibliography}{60}%
\makeatletter
\providecommand \@ifxundefined [1]{%
 \@ifx{#1\undefined}
}%
\providecommand \@ifnum [1]{%
 \ifnum #1\expandafter \@firstoftwo
 \else \expandafter \@secondoftwo
 \fi
}%
\providecommand \@ifx [1]{%
 \ifx #1\expandafter \@firstoftwo
 \else \expandafter \@secondoftwo
 \fi
}%
\providecommand \natexlab [1]{#1}%
\providecommand \enquote  [1]{``#1''}%
\providecommand \bibnamefont  [1]{#1}%
\providecommand \bibfnamefont [1]{#1}%
\providecommand \citenamefont [1]{#1}%
\providecommand \href@noop [0]{\@secondoftwo}%
\providecommand \href [0]{\begingroup \@sanitize@url \@href}%
\providecommand \@href[1]{\@@startlink{#1}\@@href}%
\providecommand \@@href[1]{\endgroup#1\@@endlink}%
\providecommand \@sanitize@url [0]{\catcode `\\12\catcode `\$12\catcode
  `\&12\catcode `\#12\catcode `\^12\catcode `\_12\catcode `\%12\relax}%
\providecommand \@@startlink[1]{}%
\providecommand \@@endlink[0]{}%
\providecommand \url  [0]{\begingroup\@sanitize@url \@url }%
\providecommand \@url [1]{\endgroup\@href {#1}{\urlprefix }}%
\providecommand \urlprefix  [0]{URL }%
\providecommand \Eprint [0]{\href }%
\providecommand \doibase [0]{https://doi.org/}%
\providecommand \selectlanguage [0]{\@gobble}%
\providecommand \bibinfo  [0]{\@secondoftwo}%
\providecommand \bibfield  [0]{\@secondoftwo}%
\providecommand \translation [1]{[#1]}%
\providecommand \BibitemOpen [0]{}%
\providecommand \bibitemStop [0]{}%
\providecommand \bibitemNoStop [0]{.\EOS\space}%
\providecommand \EOS [0]{\spacefactor3000\relax}%
\providecommand \BibitemShut  [1]{\csname bibitem#1\endcsname}%
\let\auto@bib@innerbib\@empty
\bibitem [{\citenamefont {Alberts}\ \emph {et~al.}(2008)\citenamefont
  {Alberts}, \citenamefont {Johnson}, \citenamefont {Lewis}, \citenamefont
  {Raff}, \citenamefont {Roberts},\ and\ \citenamefont {Walter}}]{AlbertsBook}%
  \BibitemOpen
  \bibfield  {author} {\bibinfo {author} {\bibfnamefont {B.}~\bibnamefont
  {Alberts}}, \bibinfo {author} {\bibfnamefont {A.}~\bibnamefont {Johnson}},
  \bibinfo {author} {\bibfnamefont {J.}~\bibnamefont {Lewis}}, \bibinfo
  {author} {\bibfnamefont {M.}~\bibnamefont {Raff}}, \bibinfo {author}
  {\bibfnamefont {K.}~\bibnamefont {Roberts}},\ and\ \bibinfo {author}
  {\bibfnamefont {P.}~\bibnamefont {Walter}},\ }\href@noop {} {\emph {\bibinfo
  {title} {{Molecular Biology of the Cell -- 5th Edition}}}}\ (\bibinfo
  {publisher} {Garland Science, Taylor and Francis Group, LLC},\ \bibinfo
  {address} {New York, NY, USA},\ \bibinfo {year} {2008})\BibitemShut {NoStop}%
\bibitem [{\citenamefont {Padhi}\ and\ \citenamefont
  {Nain}(2020)}]{Padhi2020AnnBioEng}%
  \BibitemOpen
  \bibfield  {author} {\bibinfo {author} {\bibfnamefont {A.}~\bibnamefont
  {Padhi}}\ and\ \bibinfo {author} {\bibfnamefont {A.~S.}\ \bibnamefont
  {Nain}},\ }\bibfield  {title} {\bibinfo {title} {Ecm in differentiation: A
  review of matrix structure, composition and mechanical properties},\
  }\href@noop {} {\bibfield  {journal} {\bibinfo  {journal} {Annals of
  Biomedical Engineering}\ }\textbf {\bibinfo {volume} {48}},\ \bibinfo {pages}
  {1071} (\bibinfo {year} {2020})}\BibitemShut {NoStop}%
\bibitem [{\citenamefont {Frantz}\ \emph {et~al.}(2010)\citenamefont {Frantz},
  \citenamefont {Stewart},\ and\ \citenamefont {Weaver}}]{Frantz2010JCS}%
  \BibitemOpen
  \bibfield  {author} {\bibinfo {author} {\bibfnamefont {C.}~\bibnamefont
  {Frantz}}, \bibinfo {author} {\bibfnamefont {K.~M.}\ \bibnamefont
  {Stewart}},\ and\ \bibinfo {author} {\bibfnamefont {V.~M.}\ \bibnamefont
  {Weaver}},\ }\bibfield  {title} {\bibinfo {title} {{The extracellular matrix
  at a glance}},\ }\href@noop {} {\bibfield  {journal} {\bibinfo  {journal}
  {Journal of Cell Science}\ }\textbf {\bibinfo {volume} {123}},\ \bibinfo
  {pages} {4195} (\bibinfo {year} {2010})}\BibitemShut {NoStop}%
\bibitem [{\citenamefont {Discher}\ \emph {et~al.}(2005)\citenamefont
  {Discher}, \citenamefont {Janmey},\ and\ \citenamefont
  {li~Wang}}]{Discher2005Science}%
  \BibitemOpen
  \bibfield  {author} {\bibinfo {author} {\bibfnamefont {D.~E.}\ \bibnamefont
  {Discher}}, \bibinfo {author} {\bibfnamefont {P.}~\bibnamefont {Janmey}},\
  and\ \bibinfo {author} {\bibfnamefont {Y.}~\bibnamefont {li~Wang}},\
  }\bibfield  {title} {\bibinfo {title} {Tissue cells feel and respond to the
  stiffness of their substrate},\ }\href@noop {} {\bibfield  {journal}
  {\bibinfo  {journal} {Science}\ }\textbf {\bibinfo {volume} {310}},\ \bibinfo
  {pages} {1139} (\bibinfo {year} {2005})}\BibitemShut {NoStop}%
\bibitem [{\citenamefont {Chaudhuri}\ \emph {et~al.}(2020)\citenamefont
  {Chaudhuri}, \citenamefont {Cooper-White}, \citenamefont {Janmey},
  \citenamefont {Mooney},\ and\ \citenamefont {Shenoy}}]{Chaudhuri2020Nature}%
  \BibitemOpen
  \bibfield  {author} {\bibinfo {author} {\bibfnamefont {O.}~\bibnamefont
  {Chaudhuri}}, \bibinfo {author} {\bibfnamefont {J.}~\bibnamefont
  {Cooper-White}}, \bibinfo {author} {\bibfnamefont {P.~A.}\ \bibnamefont
  {Janmey}}, \bibinfo {author} {\bibfnamefont {D.~J.}\ \bibnamefont {Mooney}},\
  and\ \bibinfo {author} {\bibfnamefont {V.~B.}\ \bibnamefont {Shenoy}},\
  }\bibfield  {title} {\bibinfo {title} {{Effects of extracellular matrix
  viscoelasticity on cellular behavior}},\ }\href@noop {} {\bibfield  {journal}
  {\bibinfo  {journal} {Nature}\ }\textbf {\bibinfo {volume} {584}},\ \bibinfo
  {pages} {535} (\bibinfo {year} {2020})}\BibitemShut {NoStop}%
\bibitem [{\citenamefont {Rozario}\ and\ \citenamefont
  {DeSimone}(2010)}]{Rozario2010DevBio}%
  \BibitemOpen
  \bibfield  {author} {\bibinfo {author} {\bibfnamefont {T.}~\bibnamefont
  {Rozario}}\ and\ \bibinfo {author} {\bibfnamefont {D.~W.}\ \bibnamefont
  {DeSimone}},\ }\bibfield  {title} {\bibinfo {title} {The extracellular matrix
  in development and morphogenesis: A dynamic view},\ }\href@noop {} {\bibfield
   {journal} {\bibinfo  {journal} {Developmental Biology}\ }\textbf {\bibinfo
  {volume} {341}},\ \bibinfo {pages} {126} (\bibinfo {year} {2010})},\ \bibinfo
  {note} {special Section: Morphogenesis}\BibitemShut {NoStop}%
\bibitem [{\citenamefont {Ulrich}\ \emph {et~al.}(2010)\citenamefont {Ulrich},
  \citenamefont {Jain}, \citenamefont {Tanner}, \citenamefont {MacKay},\ and\
  \citenamefont {Kumar}}]{Ulrich2010Biomaterials}%
  \BibitemOpen
  \bibfield  {author} {\bibinfo {author} {\bibfnamefont {T.~A.}\ \bibnamefont
  {Ulrich}}, \bibinfo {author} {\bibfnamefont {A.}~\bibnamefont {Jain}},
  \bibinfo {author} {\bibfnamefont {K.}~\bibnamefont {Tanner}}, \bibinfo
  {author} {\bibfnamefont {J.~L.}\ \bibnamefont {MacKay}},\ and\ \bibinfo
  {author} {\bibfnamefont {S.}~\bibnamefont {Kumar}},\ }\bibfield  {title}
  {\bibinfo {title} {Probing cellular mechanobiology in three-dimensional
  culture with collagen–agarose matrices},\ }\href@noop {} {\bibfield
  {journal} {\bibinfo  {journal} {Biomaterials}\ }\textbf {\bibinfo {volume}
  {31}},\ \bibinfo {pages} {1875} (\bibinfo {year} {2010})}\BibitemShut
  {NoStop}%
\bibitem [{\citenamefont {Burla}\ \emph {et~al.}(2019)\citenamefont {Burla},
  \citenamefont {Tauber}, \citenamefont {Dussi}, \citenamefont {van~der
  Gucht},\ and\ \citenamefont {Koenderink}}]{Burla2019NatPhys}%
  \BibitemOpen
  \bibfield  {author} {\bibinfo {author} {\bibfnamefont {F.}~\bibnamefont
  {Burla}}, \bibinfo {author} {\bibfnamefont {J.}~\bibnamefont {Tauber}},
  \bibinfo {author} {\bibfnamefont {S.}~\bibnamefont {Dussi}}, \bibinfo
  {author} {\bibfnamefont {J.}~\bibnamefont {van~der Gucht}},\ and\ \bibinfo
  {author} {\bibfnamefont {G.~H.}\ \bibnamefont {Koenderink}},\ }\bibfield
  {title} {\bibinfo {title} {{Stress management in composite biopolymer
  networks}},\ }\href@noop {} {\bibfield  {journal} {\bibinfo  {journal} {Nat.
  Phys.}\ }\textbf {\bibinfo {volume} {15}},\ \bibinfo {pages} {549} (\bibinfo
  {year} {2019})}\BibitemShut {NoStop}%
\bibitem [{\citenamefont {Pollard}\ and\ \citenamefont
  {Goldman}(2018)}]{Pollard2018CSHPB}%
  \BibitemOpen
  \bibfield  {author} {\bibinfo {author} {\bibfnamefont {T.~D.}\ \bibnamefont
  {Pollard}}\ and\ \bibinfo {author} {\bibfnamefont {R.~D.}\ \bibnamefont
  {Goldman}},\ }\bibfield  {title} {\bibinfo {title} {Overview of the
  cytoskeleton from an evolutionary perspective},\ }\href@noop {} {\bibfield
  {journal} {\bibinfo  {journal} {Cold Spring Harbor Perspectives in Biology}\
  }\textbf {\bibinfo {volume} {10}} (\bibinfo {year} {2018})}\BibitemShut
  {NoStop}%
\bibitem [{\citenamefont {Mostowy}\ and\ \citenamefont
  {Cossart}(2012)}]{Mostowy2012NRMCB}%
  \BibitemOpen
  \bibfield  {author} {\bibinfo {author} {\bibfnamefont {S.}~\bibnamefont
  {Mostowy}}\ and\ \bibinfo {author} {\bibfnamefont {P.}~\bibnamefont
  {Cossart}},\ }\bibfield  {title} {\bibinfo {title} {{Septins: the fourth
  component of the cytoskeleton}},\ }\href@noop {} {\bibfield  {journal}
  {\bibinfo  {journal} {Nature Reviews Molecular Cell Biology}\ }\textbf
  {\bibinfo {volume} {13}},\ \bibinfo {pages} {183} (\bibinfo {year}
  {2012})}\BibitemShut {NoStop}%
\bibitem [{\citenamefont {{Lin, Yi-Chia and Koenderink, Gijsje H. and
  MacKintosh, Frederick C. and Weitz, David A.}}(2011)}]{Lin2011SoftMatter}%
  \BibitemOpen
  \bibfield  {author} {\bibinfo {author} {\bibnamefont {{Lin, Yi-Chia and
  Koenderink, Gijsje H. and MacKintosh, Frederick C. and Weitz, David A.}}},\
  }\bibfield  {title} {\bibinfo {title} {{Control of non-linear elasticity in
  F-actin networks with microtubules}},\ }\href@noop {} {\bibfield  {journal}
  {\bibinfo  {journal} {Soft Matter}\ }\textbf {\bibinfo {volume} {7}},\
  \bibinfo {pages} {902} (\bibinfo {year} {2011})}\BibitemShut {NoStop}%
\bibitem [{\citenamefont {Jensen}\ \emph {et~al.}(2014)\citenamefont {Jensen},
  \citenamefont {Morris}, \citenamefont {Goldman},\ and\ \citenamefont
  {Weitz}}]{Jensen2014BioArch}%
  \BibitemOpen
  \bibfield  {author} {\bibinfo {author} {\bibfnamefont {M.~H.}\ \bibnamefont
  {Jensen}}, \bibinfo {author} {\bibfnamefont {E.~J.}\ \bibnamefont {Morris}},
  \bibinfo {author} {\bibfnamefont {R.~D.}\ \bibnamefont {Goldman}},\ and\
  \bibinfo {author} {\bibfnamefont {D.~A.}\ \bibnamefont {Weitz}},\ }\bibfield
  {title} {\bibinfo {title} {{Emergent properties of composite semiflexible
  biopolymer networks}},\ }\href@noop {} {\bibfield  {journal} {\bibinfo
  {journal} {BioArchitecture}\ }\textbf {\bibinfo {volume} {4}},\ \bibinfo
  {pages} {138} (\bibinfo {year} {2014})}\BibitemShut {NoStop}%
\bibitem [{\citenamefont {Gong}\ \emph {et~al.}(2003)\citenamefont {Gong},
  \citenamefont {Katsuyama}, \citenamefont {Kurokawa},\ and\ \citenamefont
  {Osada}}]{Gong2003AdvancedMat}%
  \BibitemOpen
  \bibfield  {author} {\bibinfo {author} {\bibfnamefont {J.~P.}\ \bibnamefont
  {Gong}}, \bibinfo {author} {\bibfnamefont {Y.}~\bibnamefont {Katsuyama}},
  \bibinfo {author} {\bibfnamefont {T.}~\bibnamefont {Kurokawa}},\ and\
  \bibinfo {author} {\bibfnamefont {Y.}~\bibnamefont {Osada}},\ }\bibfield
  {title} {\bibinfo {title} {Double-network hydrogels with extremely high
  mechanical strength},\ }\href@noop {} {\bibfield  {journal} {\bibinfo
  {journal} {Adv. Mater.}\ }\textbf {\bibinfo {volume} {15}},\ \bibinfo {pages}
  {1155} (\bibinfo {year} {2003})}\BibitemShut {NoStop}%
\bibitem [{\citenamefont {Gong}(2010)}]{Gong2010SoftMat}%
  \BibitemOpen
  \bibfield  {author} {\bibinfo {author} {\bibfnamefont {J.~P.}\ \bibnamefont
  {Gong}},\ }\bibfield  {title} {\bibinfo {title} {Why are double network
  hydrogels so tough?},\ }\href@noop {} {\bibfield  {journal} {\bibinfo
  {journal} {Soft Matter}\ }\textbf {\bibinfo {volume} {6}},\ \bibinfo {pages}
  {2583} (\bibinfo {year} {2010})}\BibitemShut {NoStop}%
\bibitem [{\citenamefont {Ducrot}\ \emph {et~al.}(2014)\citenamefont {Ducrot},
  \citenamefont {Chen}, \citenamefont {Bulters}, \citenamefont {Sijbesma},\
  and\ \citenamefont {Creton}}]{Ducrot2014Science}%
  \BibitemOpen
  \bibfield  {author} {\bibinfo {author} {\bibfnamefont {E.}~\bibnamefont
  {Ducrot}}, \bibinfo {author} {\bibfnamefont {Y.}~\bibnamefont {Chen}},
  \bibinfo {author} {\bibfnamefont {M.}~\bibnamefont {Bulters}}, \bibinfo
  {author} {\bibfnamefont {R.~P.}\ \bibnamefont {Sijbesma}},\ and\ \bibinfo
  {author} {\bibfnamefont {C.}~\bibnamefont {Creton}},\ }\bibfield  {title}
  {\bibinfo {title} {Toughening elastomers with sacrificial bonds and watching
  them break},\ }\href@noop {} {\bibfield  {journal} {\bibinfo  {journal}
  {Science}\ }\textbf {\bibinfo {volume} {344}},\ \bibinfo {pages} {186}
  (\bibinfo {year} {2014})}\BibitemShut {NoStop}%
\bibitem [{\citenamefont {Li}\ \emph {et~al.}(2023)\citenamefont {Li},
  \citenamefont {Cui}, \citenamefont {Zheng}, \citenamefont {Ye}, \citenamefont
  {Yu}, \citenamefont {Yang}, \citenamefont {Nakajima},\ and\ \citenamefont
  {Gong}}]{li2023SciAdv}%
  \BibitemOpen
  \bibfield  {author} {\bibinfo {author} {\bibfnamefont {X.}~\bibnamefont
  {Li}}, \bibinfo {author} {\bibfnamefont {K.}~\bibnamefont {Cui}}, \bibinfo
  {author} {\bibfnamefont {Y.}~\bibnamefont {Zheng}}, \bibinfo {author}
  {\bibfnamefont {Y.~N.}\ \bibnamefont {Ye}}, \bibinfo {author} {\bibfnamefont
  {C.}~\bibnamefont {Yu}}, \bibinfo {author} {\bibfnamefont {W.}~\bibnamefont
  {Yang}}, \bibinfo {author} {\bibfnamefont {T.}~\bibnamefont {Nakajima}},\
  and\ \bibinfo {author} {\bibfnamefont {J.~P.}\ \bibnamefont {Gong}},\
  }\bibfield  {title} {\bibinfo {title} {Role of hierarchy structure on the
  mechanical adaptation of self-healing hydrogels under cyclic stretching},\
  }\href@noop {} {\bibfield  {journal} {\bibinfo  {journal} {Sci. Adv.}\
  }\textbf {\bibinfo {volume} {9}},\ \bibinfo {pages} {eadj6856} (\bibinfo
  {year} {2023})}\BibitemShut {NoStop}%
\bibitem [{\citenamefont {Yu}\ \emph {et~al.}(2020)\citenamefont {Yu},
  \citenamefont {Zheng}, \citenamefont {Fang}, \citenamefont {Ying},
  \citenamefont {Du}, \citenamefont {Wang}, \citenamefont {Ren}, \citenamefont
  {Wu},\ and\ \citenamefont {Zheng}}]{yu2020AdvMater}%
  \BibitemOpen
  \bibfield  {author} {\bibinfo {author} {\bibfnamefont {H.~C.}\ \bibnamefont
  {Yu}}, \bibinfo {author} {\bibfnamefont {S.~Y.}\ \bibnamefont {Zheng}},
  \bibinfo {author} {\bibfnamefont {L.}~\bibnamefont {Fang}}, \bibinfo {author}
  {\bibfnamefont {Z.}~\bibnamefont {Ying}}, \bibinfo {author} {\bibfnamefont
  {M.}~\bibnamefont {Du}}, \bibinfo {author} {\bibfnamefont {J.}~\bibnamefont
  {Wang}}, \bibinfo {author} {\bibfnamefont {K.-F.}\ \bibnamefont {Ren}},
  \bibinfo {author} {\bibfnamefont {Z.~L.}\ \bibnamefont {Wu}},\ and\ \bibinfo
  {author} {\bibfnamefont {Q.}~\bibnamefont {Zheng}},\ }\bibfield  {title}
  {\bibinfo {title} {Reversibly transforming a highly swollen polyelectrolyte
  hydrogel to an extremely tough one and its application as a tubular
  grasper},\ }\href@noop {} {\bibfield  {journal} {\bibinfo  {journal} {Adv.
  Mater.}\ }\textbf {\bibinfo {volume} {32}},\ \bibinfo {pages} {2005171}
  (\bibinfo {year} {2020})}\BibitemShut {NoStop}%
\bibitem [{\citenamefont {Li}\ and\ \citenamefont
  {Gong}(2024)}]{li2024dnatrevmat}%
  \BibitemOpen
  \bibfield  {author} {\bibinfo {author} {\bibfnamefont {X.}~\bibnamefont
  {Li}}\ and\ \bibinfo {author} {\bibfnamefont {J.~P.}\ \bibnamefont {Gong}},\
  }\bibfield  {title} {\bibinfo {title} {Design principles for strong and tough
  hydrogels},\ }\href@noop {} {\bibfield  {journal} {\bibinfo  {journal}
  {Nature Reviews Materials}\ ,\ \bibinfo {pages} {1}} (\bibinfo {year}
  {2024})}\BibitemShut {NoStop}%
\bibitem [{\citenamefont {Li}\ \emph {et~al.}(2020)\citenamefont {Li},
  \citenamefont {Cui}, \citenamefont {Sun}, \citenamefont {Meng}, \citenamefont
  {Yu}, \citenamefont {Li}, \citenamefont {Creton}, \citenamefont {Kurokawa},\
  and\ \citenamefont {Gong}}]{li2020PNAS}%
  \BibitemOpen
  \bibfield  {author} {\bibinfo {author} {\bibfnamefont {X.}~\bibnamefont
  {Li}}, \bibinfo {author} {\bibfnamefont {K.}~\bibnamefont {Cui}}, \bibinfo
  {author} {\bibfnamefont {T.~L.}\ \bibnamefont {Sun}}, \bibinfo {author}
  {\bibfnamefont {L.}~\bibnamefont {Meng}}, \bibinfo {author} {\bibfnamefont
  {C.}~\bibnamefont {Yu}}, \bibinfo {author} {\bibfnamefont {L.}~\bibnamefont
  {Li}}, \bibinfo {author} {\bibfnamefont {C.}~\bibnamefont {Creton}}, \bibinfo
  {author} {\bibfnamefont {T.}~\bibnamefont {Kurokawa}},\ and\ \bibinfo
  {author} {\bibfnamefont {J.~P.}\ \bibnamefont {Gong}},\ }\bibfield  {title}
  {\bibinfo {title} {Mesoscale bicontinuous networks in self-healing hydrogels
  delay fatigue fracture},\ }\href@noop {} {\bibfield  {journal} {\bibinfo
  {journal} {PNAS}\ }\textbf {\bibinfo {volume} {117}},\ \bibinfo {pages}
  {7606} (\bibinfo {year} {2020})}\BibitemShut {NoStop}%
\bibitem [{\citenamefont {Webber}\ \emph {et~al.}(2007)\citenamefont {Webber},
  \citenamefont {Creton}, \citenamefont {Brown},\ and\ \citenamefont
  {Gong}}]{webber2007Macro}%
  \BibitemOpen
  \bibfield  {author} {\bibinfo {author} {\bibfnamefont {R.~E.}\ \bibnamefont
  {Webber}}, \bibinfo {author} {\bibfnamefont {C.}~\bibnamefont {Creton}},
  \bibinfo {author} {\bibfnamefont {H.~R.}\ \bibnamefont {Brown}},\ and\
  \bibinfo {author} {\bibfnamefont {J.~P.}\ \bibnamefont {Gong}},\ }\bibfield
  {title} {\bibinfo {title} {Large strain hysteresis and mullins effect of
  tough double-network hydrogels},\ }\href@noop {} {\bibfield  {journal}
  {\bibinfo  {journal} {Macromolecules}\ }\textbf {\bibinfo {volume} {40}},\
  \bibinfo {pages} {2919} (\bibinfo {year} {2007})}\BibitemShut {NoStop}%
\bibitem [{\citenamefont {Glotzer}\ and\ \citenamefont
  {Solomon}(2007)}]{glotzer2007anisotropy}%
  \BibitemOpen
  \bibfield  {author} {\bibinfo {author} {\bibfnamefont {S.}~\bibnamefont
  {Glotzer}}\ and\ \bibinfo {author} {\bibfnamefont {M.}~\bibnamefont
  {Solomon}},\ }\bibfield  {title} {\bibinfo {title} {Anisotropy of building
  blocks and their assembly into complex structures},\ }\href@noop {}
  {\bibfield  {journal} {\bibinfo  {journal} {Nature Materials}\ }\textbf
  {\bibinfo {volume} {6}},\ \bibinfo {pages} {557} (\bibinfo {year}
  {2007})}\BibitemShut {NoStop}%
\bibitem [{\citenamefont {Grzelczak}\ \emph {et~al.}(2010)\citenamefont
  {Grzelczak}, \citenamefont {Vermant}, \citenamefont {Furst},\ and\
  \citenamefont {Liz-Marzán}}]{Grzelczak2010directed}%
  \BibitemOpen
  \bibfield  {author} {\bibinfo {author} {\bibfnamefont {M.}~\bibnamefont
  {Grzelczak}}, \bibinfo {author} {\bibfnamefont {J.}~\bibnamefont {Vermant}},
  \bibinfo {author} {\bibfnamefont {E.~M.}\ \bibnamefont {Furst}},\ and\
  \bibinfo {author} {\bibfnamefont {L.~M.}\ \bibnamefont {Liz-Marzán}},\
  }\bibfield  {title} {\bibinfo {title} {Directed self-assembly of
  nanoparticles},\ }\href@noop {} {\bibfield  {journal} {\bibinfo  {journal}
  {ACS Nano}\ }\textbf {\bibinfo {volume} {4}},\ \bibinfo {pages} {3591}
  (\bibinfo {year} {2010})}\BibitemShut {NoStop}%
\bibitem [{\citenamefont {Chen}\ \emph {et~al.}(2011)\citenamefont {Chen},
  \citenamefont {Diesel}, \citenamefont {Whitmer}, \citenamefont {Bae},
  \citenamefont {Luijten},\ and\ \citenamefont {Granick}}]{Chen2011triblock}%
  \BibitemOpen
  \bibfield  {author} {\bibinfo {author} {\bibfnamefont {Q.}~\bibnamefont
  {Chen}}, \bibinfo {author} {\bibfnamefont {E.}~\bibnamefont {Diesel}},
  \bibinfo {author} {\bibfnamefont {J.~K.}\ \bibnamefont {Whitmer}}, \bibinfo
  {author} {\bibfnamefont {S.~C.}\ \bibnamefont {Bae}}, \bibinfo {author}
  {\bibfnamefont {E.}~\bibnamefont {Luijten}},\ and\ \bibinfo {author}
  {\bibfnamefont {S.}~\bibnamefont {Granick}},\ }\bibfield  {title} {\bibinfo
  {title} {Triblock colloids for directed self-assembly},\ }\href@noop {}
  {\bibfield  {journal} {\bibinfo  {journal} {JACS}\ }\textbf {\bibinfo
  {volume} {133}},\ \bibinfo {pages} {7725} (\bibinfo {year}
  {2011})}\BibitemShut {NoStop}%
\bibitem [{\citenamefont {Aida}\ \emph {et~al.}(2012)\citenamefont {Aida},
  \citenamefont {Mejier},\ and\ \citenamefont {Stupp}}]{Aida2012functional}%
  \BibitemOpen
  \bibfield  {author} {\bibinfo {author} {\bibfnamefont {T.}~\bibnamefont
  {Aida}}, \bibinfo {author} {\bibfnamefont {E.}~\bibnamefont {Mejier}},\ and\
  \bibinfo {author} {\bibfnamefont {S.~I.}\ \bibnamefont {Stupp}},\ }\bibfield
  {title} {\bibinfo {title} {Functional supramolecular polymers},\ }\href@noop
  {} {\bibfield  {journal} {\bibinfo  {journal} {Science}\ }\textbf {\bibinfo
  {volume} {335}},\ \bibinfo {pages} {813} (\bibinfo {year}
  {2012})}\BibitemShut {NoStop}%
\bibitem [{\citenamefont {Kotov}\ and\ \citenamefont
  {Weiss}(2014)}]{Kotov2014self}%
  \BibitemOpen
  \bibfield  {author} {\bibinfo {author} {\bibfnamefont {N.~A.}\ \bibnamefont
  {Kotov}}\ and\ \bibinfo {author} {\bibfnamefont {P.~S.}\ \bibnamefont
  {Weiss}},\ }\bibfield  {title} {\bibinfo {title} {Self-assembly of
  nanoparticles: A snapshot},\ }\href@noop {} {\bibfield  {journal} {\bibinfo
  {journal} {ACS Nano}\ }\textbf {\bibinfo {volume} {8}},\ \bibinfo {pages}
  {3101} (\bibinfo {year} {2014})}\BibitemShut {NoStop}%
\bibitem [{\citenamefont {Travesset}(2015)}]{travesset2015binary}%
  \BibitemOpen
  \bibfield  {author} {\bibinfo {author} {\bibfnamefont {A.}~\bibnamefont
  {Travesset}},\ }\bibfield  {title} {\bibinfo {title} {Binary nanoparticle
  superlattices of soft-particle systems},\ }\href@noop {} {\bibfield
  {journal} {\bibinfo  {journal} {PNAS}\ }\textbf {\bibinfo {volume} {112}},\
  \bibinfo {pages} {9563} (\bibinfo {year} {2015})}\BibitemShut {NoStop}%
\bibitem [{\citenamefont {Varrato}\ \emph {et~al.}(2012)\citenamefont
  {Varrato}, \citenamefont {Michele}, \citenamefont {Belushkin}, \citenamefont
  {Dorsaz}, \citenamefont {Nathan}, \citenamefont {Eiser},\ and\ \citenamefont
  {Foffi}}]{Varrato2012PNAS}%
  \BibitemOpen
  \bibfield  {author} {\bibinfo {author} {\bibfnamefont {F.}~\bibnamefont
  {Varrato}}, \bibinfo {author} {\bibfnamefont {L.~D.}\ \bibnamefont
  {Michele}}, \bibinfo {author} {\bibfnamefont {M.}~\bibnamefont {Belushkin}},
  \bibinfo {author} {\bibfnamefont {N.}~\bibnamefont {Dorsaz}}, \bibinfo
  {author} {\bibfnamefont {S.~H.}\ \bibnamefont {Nathan}}, \bibinfo {author}
  {\bibfnamefont {E.}~\bibnamefont {Eiser}},\ and\ \bibinfo {author}
  {\bibfnamefont {G.}~\bibnamefont {Foffi}},\ }\bibfield  {title} {\bibinfo
  {title} {Arrested demixing opens route to bigels},\ }\href@noop {} {\bibfield
   {journal} {\bibinfo  {journal} {Proceedings of the National Academy of
  Sciences}\ }\textbf {\bibinfo {volume} {109}},\ \bibinfo {pages} {19155}
  (\bibinfo {year} {2012})}\BibitemShut {NoStop}%
\bibitem [{\citenamefont {Immink}\ \emph {et~al.}(2019)\citenamefont {Immink},
  \citenamefont {Maris}, \citenamefont {Crassous}, \citenamefont {Stenhammar},\
  and\ \citenamefont {Schurtenberger}}]{Immink2019reversible}%
  \BibitemOpen
  \bibfield  {author} {\bibinfo {author} {\bibfnamefont {J.~N.}\ \bibnamefont
  {Immink}}, \bibinfo {author} {\bibfnamefont {J.~J.~E.}\ \bibnamefont
  {Maris}}, \bibinfo {author} {\bibfnamefont {J.~J.}\ \bibnamefont {Crassous}},
  \bibinfo {author} {\bibfnamefont {J.}~\bibnamefont {Stenhammar}},\ and\
  \bibinfo {author} {\bibfnamefont {P.}~\bibnamefont {Schurtenberger}},\
  }\bibfield  {title} {\bibinfo {title} {Reversible formation of
  thermoresponsive binary particle gels with tunable structural and mechanical
  properties},\ }\href@noop {} {\bibfield  {journal} {\bibinfo  {journal} {ACS
  Nano}\ }\textbf {\bibinfo {volume} {13}},\ \bibinfo {pages} {3292} (\bibinfo
  {year} {2019})}\BibitemShut {NoStop}%
\bibitem [{\citenamefont {Rijns}\ \emph {et~al.}(2024)\citenamefont {Rijns},
  \citenamefont {Rutten}, \citenamefont {Bellan}, \citenamefont {Yuang},
  \citenamefont {Mugnai}, \citenamefont {Rocha}, \citenamefont {del Gado},
  \citenamefont {Kouwer},\ and\ \citenamefont {Dankers}}]{Rijns2024}%
  \BibitemOpen
  \bibfield  {author} {\bibinfo {author} {\bibfnamefont {L.}~\bibnamefont
  {Rijns}}, \bibinfo {author} {\bibfnamefont {M.~G.}\ \bibnamefont {Rutten}},
  \bibinfo {author} {\bibfnamefont {R.}~\bibnamefont {Bellan}}, \bibinfo
  {author} {\bibfnamefont {H.}~\bibnamefont {Yuang}}, \bibinfo {author}
  {\bibfnamefont {M.}~\bibnamefont {Mugnai}}, \bibinfo {author} {\bibfnamefont
  {S.}~\bibnamefont {Rocha}}, \bibinfo {author} {\bibfnamefont
  {E.}~\bibnamefont {del Gado}}, \bibinfo {author} {\bibfnamefont {P.~H.~J.}\
  \bibnamefont {Kouwer}},\ and\ \bibinfo {author} {\bibfnamefont {P.~Y.~W.}\
  \bibnamefont {Dankers}},\ }\bibfield  {title} {\bibinfo {title} {Synthetic,
  multi-dynamic hydrogels by uniting stress stiffening and supramolecular
  polymers},\ }\href@noop {} {\bibfield  {journal} {\bibinfo  {journal}
  {preprint}\ } (\bibinfo {year} {2024})}\BibitemShut {NoStop}%
\bibitem [{\citenamefont {Kim}\ \emph {et~al.}(2024)\citenamefont {Kim},
  \citenamefont {Akkunuri}, \citenamefont {Qian}, \citenamefont {Yao},
  \citenamefont {Sun}, \citenamefont {Chen}, \citenamefont {Vo},\ and\
  \citenamefont {Chen}}]{Kim2024direct}%
  \BibitemOpen
  \bibfield  {author} {\bibinfo {author} {\bibfnamefont {A.}~\bibnamefont
  {Kim}}, \bibinfo {author} {\bibfnamefont {K.}~\bibnamefont {Akkunuri}},
  \bibinfo {author} {\bibfnamefont {C.}~\bibnamefont {Qian}}, \bibinfo {author}
  {\bibfnamefont {L.}~\bibnamefont {Yao}}, \bibinfo {author} {\bibfnamefont
  {K.}~\bibnamefont {Sun}}, \bibinfo {author} {\bibfnamefont {Z.}~\bibnamefont
  {Chen}}, \bibinfo {author} {\bibfnamefont {T.}~\bibnamefont {Vo}},\ and\
  \bibinfo {author} {\bibfnamefont {Q.}~\bibnamefont {Chen}},\ }\bibfield
  {title} {\bibinfo {title} {Direct imaging of “patch-clasping” and
  relaxation in robust and flexible nanoparticle assemblies},\ }\href@noop {}
  {\bibfield  {journal} {\bibinfo  {journal} {ACS Nano}\ }\textbf {\bibinfo
  {volume} {18}},\ \bibinfo {pages} {939} (\bibinfo {year} {2024})}\BibitemShut
  {NoStop}%
\bibitem [{\citenamefont {Mao}\ and\ \citenamefont {Kotov}(2024)}]{Mao2024MRS}%
  \BibitemOpen
  \bibfield  {author} {\bibinfo {author} {\bibfnamefont {X.}~\bibnamefont
  {Mao}}\ and\ \bibinfo {author} {\bibfnamefont {N.}~\bibnamefont {Kotov}},\
  }\bibfield  {title} {\bibinfo {title} {Complexity, disorder, and
  functionality of nanoscale materials},\ }\href@noop {} {\bibfield  {journal}
  {\bibinfo  {journal} {MRS Bulletin}\ }\textbf {\bibinfo {volume} {49}},\
  \bibinfo {pages} {352–364} (\bibinfo {year} {2024})}\BibitemShut {NoStop}%
\bibitem [{\citenamefont {Fung}(2013)}]{fung2013biomechanics}%
  \BibitemOpen
  \bibfield  {author} {\bibinfo {author} {\bibfnamefont {Y.-c.}\ \bibnamefont
  {Fung}},\ }\href@noop {} {\emph {\bibinfo {title} {Biomechanics: mechanical
  properties of living tissues}}}\ (\bibinfo  {publisher} {Springer Science \&
  Business Media},\ \bibinfo {year} {2013})\BibitemShut {NoStop}%
\bibitem [{\citenamefont {Moutos}\ \emph {et~al.}(2007)\citenamefont {Moutos},
  \citenamefont {Freed},\ and\ \citenamefont {Guilak}}]{moutos2007nautureM}%
  \BibitemOpen
  \bibfield  {author} {\bibinfo {author} {\bibfnamefont {F.~T.}\ \bibnamefont
  {Moutos}}, \bibinfo {author} {\bibfnamefont {L.~E.}\ \bibnamefont {Freed}},\
  and\ \bibinfo {author} {\bibfnamefont {F.}~\bibnamefont {Guilak}},\
  }\bibfield  {title} {\bibinfo {title} {A biomimetic three-dimensional woven
  composite scaffold for functional tissue engineering of cartilage},\
  }\href@noop {} {\bibfield  {journal} {\bibinfo  {journal} {Nat. Mater}\
  }\textbf {\bibinfo {volume} {6}},\ \bibinfo {pages} {162} (\bibinfo {year}
  {2007})}\BibitemShut {NoStop}%
\bibitem [{\citenamefont {Lee}\ and\ \citenamefont
  {Mooney}(2001)}]{lee2001ChemRev}%
  \BibitemOpen
  \bibfield  {author} {\bibinfo {author} {\bibfnamefont {K.~Y.}\ \bibnamefont
  {Lee}}\ and\ \bibinfo {author} {\bibfnamefont {D.~J.}\ \bibnamefont
  {Mooney}},\ }\bibfield  {title} {\bibinfo {title} {Hydrogels for tissue
  engineering},\ }\href@noop {} {\bibfield  {journal} {\bibinfo  {journal}
  {Chem. Rev.}\ }\textbf {\bibinfo {volume} {101}},\ \bibinfo {pages} {1869}
  (\bibinfo {year} {2001})}\BibitemShut {NoStop}%
\bibitem [{\citenamefont {Wallin}\ \emph {et~al.}(2020)\citenamefont {Wallin},
  \citenamefont {Simonsen}, \citenamefont {Pan}, \citenamefont {Wang},
  \citenamefont {Giannelis}, \citenamefont {Shepherd},\ and\ \citenamefont
  {Mengüç}}]{Wallin2020NatComm}%
  \BibitemOpen
  \bibfield  {author} {\bibinfo {author} {\bibfnamefont {T.~J.}\ \bibnamefont
  {Wallin}}, \bibinfo {author} {\bibfnamefont {L.-E.}\ \bibnamefont
  {Simonsen}}, \bibinfo {author} {\bibfnamefont {W.}~\bibnamefont {Pan}},
  \bibinfo {author} {\bibfnamefont {K.}~\bibnamefont {Wang}}, \bibinfo {author}
  {\bibfnamefont {E.}~\bibnamefont {Giannelis}}, \bibinfo {author}
  {\bibfnamefont {R.~F.}\ \bibnamefont {Shepherd}},\ and\ \bibinfo {author}
  {\bibfnamefont {Y.}~\bibnamefont {Mengüç}},\ }\bibfield  {title} {\bibinfo
  {title} {3d printable tough silicone double networks},\ }\href@noop {}
  {\bibfield  {journal} {\bibinfo  {journal} {Nature Communications}\ }\textbf
  {\bibinfo {volume} {11}},\ \bibinfo {pages} {4000} (\bibinfo {year}
  {2020})}\BibitemShut {NoStop}%
\bibitem [{\citenamefont {Romberg}\ and\ \citenamefont
  {Kotula}(2023)}]{Romberg2023AdditMan}%
  \BibitemOpen
  \bibfield  {author} {\bibinfo {author} {\bibfnamefont {S.~K.}\ \bibnamefont
  {Romberg}}\ and\ \bibinfo {author} {\bibfnamefont {A.~P.}\ \bibnamefont
  {Kotula}},\ }\bibfield  {title} {\bibinfo {title} {Simultaneous rheology and
  cure kinetics dictate thermal post-curing of thermoset composite resins for
  material extrusion},\ }\href@noop {} {\bibfield  {journal} {\bibinfo
  {journal} {Additive Manufacturing}\ }\textbf {\bibinfo {volume} {71}},\
  \bibinfo {pages} {103589} (\bibinfo {year} {2023})}\BibitemShut {NoStop}%
\bibitem [{\citenamefont {Ferreiro-C\'{o}rdova}\ \emph
  {et~al.}(2020)\citenamefont {Ferreiro-C\'{o}rdova}, \citenamefont {Del~Gado},
  \citenamefont {Foffi},\ and\ \citenamefont
  {Bouzid}}]{FerreiroCordova2020SoftMatter}%
  \BibitemOpen
  \bibfield  {author} {\bibinfo {author} {\bibfnamefont {C.}~\bibnamefont
  {Ferreiro-C\'{o}rdova}}, \bibinfo {author} {\bibfnamefont {E.}~\bibnamefont
  {Del~Gado}}, \bibinfo {author} {\bibfnamefont {G.}~\bibnamefont {Foffi}},\
  and\ \bibinfo {author} {\bibfnamefont {M.}~\bibnamefont {Bouzid}},\
  }\bibfield  {title} {\bibinfo {title} {Multi-component colloidal gels:
  interplay between structure and mechanical properties},\ }\href@noop {}
  {\bibfield  {journal} {\bibinfo  {journal} {Soft Matter}\ }\textbf {\bibinfo
  {volume} {16}},\ \bibinfo {pages} {4414} (\bibinfo {year}
  {2020})}\BibitemShut {NoStop}%
\bibitem [{\citenamefont {Bantawa}\ \emph {et~al.}(2021)\citenamefont
  {Bantawa}, \citenamefont {Fontaine-Seiler}, \citenamefont {Olmsted},\ and\
  \citenamefont {Del~Gado}}]{Bantawa2021JPhysCondMat}%
  \BibitemOpen
  \bibfield  {author} {\bibinfo {author} {\bibfnamefont {M.}~\bibnamefont
  {Bantawa}}, \bibinfo {author} {\bibfnamefont {W.~A.}\ \bibnamefont
  {Fontaine-Seiler}}, \bibinfo {author} {\bibfnamefont {P.~D.}\ \bibnamefont
  {Olmsted}},\ and\ \bibinfo {author} {\bibfnamefont {E.}~\bibnamefont
  {Del~Gado}},\ }\bibfield  {title} {\bibinfo {title} {{Microscopic
  interactions and emerging elasticity in model soft particulate gels}},\
  }\href@noop {} {\bibfield  {journal} {\bibinfo  {journal} {Journal of
  Physics: Condensed Matter}\ }\textbf {\bibinfo {volume} {33}},\ \bibinfo
  {pages} {414001} (\bibinfo {year} {2021})}\BibitemShut {NoStop}%
\bibitem [{\citenamefont {Colombo}\ \emph {et~al.}(2013)\citenamefont
  {Colombo}, \citenamefont {Widmer-Cooper},\ and\ \citenamefont
  {Del~Gado}}]{Colombo2013PRL}%
  \BibitemOpen
  \bibfield  {author} {\bibinfo {author} {\bibfnamefont {J.}~\bibnamefont
  {Colombo}}, \bibinfo {author} {\bibfnamefont {A.}~\bibnamefont
  {Widmer-Cooper}},\ and\ \bibinfo {author} {\bibfnamefont {E.}~\bibnamefont
  {Del~Gado}},\ }\bibfield  {title} {\bibinfo {title} {{Microscopic Picture of
  Cooperative Processes in Restructuring Gel Networks}},\ }\href@noop {}
  {\bibfield  {journal} {\bibinfo  {journal} {Phys. Rev. Lett.}\ }\textbf
  {\bibinfo {volume} {110}},\ \bibinfo {pages} {198301} (\bibinfo {year}
  {2013})}\BibitemShut {NoStop}%
\bibitem [{\citenamefont {Colombo}\ and\ \citenamefont
  {Del~Gado}(2014{\natexlab{a}})}]{Colombo2014SoftMatter}%
  \BibitemOpen
  \bibfield  {author} {\bibinfo {author} {\bibfnamefont {J.}~\bibnamefont
  {Colombo}}\ and\ \bibinfo {author} {\bibfnamefont {E.}~\bibnamefont
  {Del~Gado}},\ }\bibfield  {title} {\bibinfo {title} {{Self-assembly and
  cooperative dynamics of a model colloidal gel network}},\ }\href@noop {}
  {\bibfield  {journal} {\bibinfo  {journal} {Soft Matter}\ }\textbf {\bibinfo
  {volume} {10}},\ \bibinfo {pages} {4003} (\bibinfo {year}
  {2014}{\natexlab{a}})}\BibitemShut {NoStop}%
\bibitem [{\citenamefont {Colombo}\ and\ \citenamefont
  {Del~Gado}(2014{\natexlab{b}})}]{Colombo2014JofRheology}%
  \BibitemOpen
  \bibfield  {author} {\bibinfo {author} {\bibfnamefont {J.}~\bibnamefont
  {Colombo}}\ and\ \bibinfo {author} {\bibfnamefont {E.}~\bibnamefont
  {Del~Gado}},\ }\bibfield  {title} {\bibinfo {title} {Stress localization,
  stiffening, and yielding in a model colloidal gel},\ }\href@noop {}
  {\bibfield  {journal} {\bibinfo  {journal} {J. Rheol.}\ }\textbf {\bibinfo
  {volume} {58}},\ \bibinfo {pages} {1089} (\bibinfo {year}
  {2014}{\natexlab{b}})}\BibitemShut {NoStop}%
\bibitem [{\citenamefont {Bouzid}\ \emph {et~al.}(2017)\citenamefont {Bouzid},
  \citenamefont {Colombo}, \citenamefont {Barbosa},\ and\ \citenamefont
  {Del~Gado}}]{Bouzid2017NatComm}%
  \BibitemOpen
  \bibfield  {author} {\bibinfo {author} {\bibfnamefont {M.}~\bibnamefont
  {Bouzid}}, \bibinfo {author} {\bibfnamefont {J.}~\bibnamefont {Colombo}},
  \bibinfo {author} {\bibfnamefont {L.~V.}\ \bibnamefont {Barbosa}},\ and\
  \bibinfo {author} {\bibfnamefont {E.}~\bibnamefont {Del~Gado}},\ }\bibfield
  {title} {\bibinfo {title} {{Elastically driven intermittent microscopic
  dynamics in soft solids}},\ }\href@noop {} {\bibfield  {journal} {\bibinfo
  {journal} {Nat. Commun.}\ }\textbf {\bibinfo {volume} {8}},\ \bibinfo {pages}
  {15846} (\bibinfo {year} {2017})}\BibitemShut {NoStop}%
\bibitem [{\citenamefont {Bouzid}\ and\ \citenamefont
  {Del~Gado}(2018)}]{Bouzid2018Langmuir}%
  \BibitemOpen
  \bibfield  {author} {\bibinfo {author} {\bibfnamefont {M.}~\bibnamefont
  {Bouzid}}\ and\ \bibinfo {author} {\bibfnamefont {E.}~\bibnamefont
  {Del~Gado}},\ }\bibfield  {title} {\bibinfo {title} {{Network Topology in
  Soft Gels: Hardening and Softening Materials}},\ }\href@noop {} {\bibfield
  {journal} {\bibinfo  {journal} {Langmuir}\ }\textbf {\bibinfo {volume}
  {34}},\ \bibinfo {pages} {773} (\bibinfo {year} {2018})}\BibitemShut
  {NoStop}%
\bibitem [{\citenamefont {Bantawa}\ \emph {et~al.}(2023)\citenamefont
  {Bantawa}, \citenamefont {Keshavarz}, \citenamefont {Geri}, \citenamefont
  {Bouzid}, \citenamefont {Divoux}, \citenamefont {McKinley},\ and\
  \citenamefont {Del~Gado}}]{Bantawa2023hidden}%
  \BibitemOpen
  \bibfield  {author} {\bibinfo {author} {\bibfnamefont {M.}~\bibnamefont
  {Bantawa}}, \bibinfo {author} {\bibfnamefont {B.}~\bibnamefont {Keshavarz}},
  \bibinfo {author} {\bibfnamefont {M.}~\bibnamefont {Geri}}, \bibinfo {author}
  {\bibfnamefont {M.}~\bibnamefont {Bouzid}}, \bibinfo {author} {\bibfnamefont
  {T.}~\bibnamefont {Divoux}}, \bibinfo {author} {\bibfnamefont
  {G.}~\bibnamefont {McKinley}},\ and\ \bibinfo {author} {\bibfnamefont
  {E.}~\bibnamefont {Del~Gado}},\ }\bibfield  {title} {\bibinfo {title} {The
  hidden hierachical nature of soft particulate gels},\ }\href@noop {}
  {\bibfield  {journal} {\bibinfo  {journal} {Nature Phys.}\ }\textbf {\bibinfo
  {volume} {19}},\ \bibinfo {pages} {1178} (\bibinfo {year}
  {2023})}\BibitemShut {NoStop}%
\bibitem [{\citenamefont {Vereroudakis}\ \emph {et~al.}(2020)\citenamefont
  {Vereroudakis}, \citenamefont {Bantawa}, \citenamefont {Lafleur},
  \citenamefont {Parisi}, \citenamefont {Matsumoto}, \citenamefont {Peeters},
  \citenamefont {Del~Gado}, \citenamefont {Meijer},\ and\ \citenamefont
  {Vlassopoulos}}]{Vereroudakis2020ACSCS}%
  \BibitemOpen
  \bibfield  {author} {\bibinfo {author} {\bibfnamefont {E.}~\bibnamefont
  {Vereroudakis}}, \bibinfo {author} {\bibfnamefont {M.}~\bibnamefont
  {Bantawa}}, \bibinfo {author} {\bibfnamefont {R.~P.~M.}\ \bibnamefont
  {Lafleur}}, \bibinfo {author} {\bibfnamefont {D.}~\bibnamefont {Parisi}},
  \bibinfo {author} {\bibfnamefont {N.~M.}\ \bibnamefont {Matsumoto}}, \bibinfo
  {author} {\bibfnamefont {J.~W.}\ \bibnamefont {Peeters}}, \bibinfo {author}
  {\bibfnamefont {E.}~\bibnamefont {Del~Gado}}, \bibinfo {author}
  {\bibfnamefont {E.~W.}\ \bibnamefont {Meijer}},\ and\ \bibinfo {author}
  {\bibfnamefont {D.}~\bibnamefont {Vlassopoulos}},\ }\bibfield  {title}
  {\bibinfo {title} {{Competitive Supramolecular Associations Mediate the
  Viscoelasticity of Binary Hydrogels}},\ }\href@noop {} {\bibfield  {journal}
  {\bibinfo  {journal} {ACS Cent Sci}\ }\textbf {\bibinfo {volume} {6}},\
  \bibinfo {pages} {1401} (\bibinfo {year} {2020})}\BibitemShut {NoStop}%
\bibitem [{\citenamefont {Donley}\ \emph {et~al.}(2022)\citenamefont {Donley},
  \citenamefont {Bantawa},\ and\ \citenamefont
  {Del~Gado}}]{Donley2022JofRheol}%
  \BibitemOpen
  \bibfield  {author} {\bibinfo {author} {\bibfnamefont {G.~J.}\ \bibnamefont
  {Donley}}, \bibinfo {author} {\bibfnamefont {M.}~\bibnamefont {Bantawa}},\
  and\ \bibinfo {author} {\bibfnamefont {E.}~\bibnamefont {Del~Gado}},\
  }\bibfield  {title} {\bibinfo {title} {{Time-resolved microstructural changes
  in large amplitude oscillatory shear of model single and double component
  soft gels}},\ }\href@noop {} {\bibfield  {journal} {\bibinfo  {journal}
  {Journal of Rheology}\ }\textbf {\bibinfo {volume} {66}},\ \bibinfo {pages}
  {1287} (\bibinfo {year} {2022})}\BibitemShut {NoStop}%
\bibitem [{\citenamefont {Dellatolas}\ \emph {et~al.}(2023)\citenamefont
  {Dellatolas}, \citenamefont {Bantawa}, \citenamefont {Damerau}, \citenamefont
  {Guo}, \citenamefont {Divoux}, \citenamefont {Del~Gado},\ and\ \citenamefont
  {Bischofberger}}]{Dellatolas2023local}%
  \BibitemOpen
  \bibfield  {author} {\bibinfo {author} {\bibfnamefont {I.}~\bibnamefont
  {Dellatolas}}, \bibinfo {author} {\bibfnamefont {M.}~\bibnamefont {Bantawa}},
  \bibinfo {author} {\bibfnamefont {B.}~\bibnamefont {Damerau}}, \bibinfo
  {author} {\bibfnamefont {M.}~\bibnamefont {Guo}}, \bibinfo {author}
  {\bibfnamefont {T.}~\bibnamefont {Divoux}}, \bibinfo {author} {\bibfnamefont
  {E.}~\bibnamefont {Del~Gado}},\ and\ \bibinfo {author} {\bibfnamefont
  {I.}~\bibnamefont {Bischofberger}},\ }\bibfield  {title} {\bibinfo {title}
  {Local mechanism governs global reinforcement of nanofiller-hydrogel
  composites},\ }\href@noop {} {\bibfield  {journal} {\bibinfo  {journal} {ACS
  Nano}\ }\textbf {\bibinfo {volume} {17}},\ \bibinfo {pages} {20939} (\bibinfo
  {year} {2023})}\BibitemShut {NoStop}%
\bibitem [{\citenamefont {Hill}(1986)}]{HillBook}%
  \BibitemOpen
  \bibfield  {author} {\bibinfo {author} {\bibfnamefont {T.~L.}\ \bibnamefont
  {Hill}},\ }\href@noop {} {\emph {\bibinfo {title} {An Introduction to
  Statistical Thermodynamics}}}\ (\bibinfo  {publisher} {Dover Publications},\
  \bibinfo {year} {1986})\BibitemShut {NoStop}%
\bibitem [{\citenamefont {Geri}\ \emph {et~al.}(2018)\citenamefont {Geri},
  \citenamefont {Keshavarz}, \citenamefont {Divoux}, \citenamefont {Clasen},
  \citenamefont {Curtis},\ and\ \citenamefont {McKinley}}]{Geri2018PRX}%
  \BibitemOpen
  \bibfield  {author} {\bibinfo {author} {\bibfnamefont {M.}~\bibnamefont
  {Geri}}, \bibinfo {author} {\bibfnamefont {B.}~\bibnamefont {Keshavarz}},
  \bibinfo {author} {\bibfnamefont {T.}~\bibnamefont {Divoux}}, \bibinfo
  {author} {\bibfnamefont {C.}~\bibnamefont {Clasen}}, \bibinfo {author}
  {\bibfnamefont {D.~J.}\ \bibnamefont {Curtis}},\ and\ \bibinfo {author}
  {\bibfnamefont {G.~H.}\ \bibnamefont {McKinley}},\ }\bibfield  {title}
  {\bibinfo {title} {Time-resolved mechanical spectroscopy of soft materials
  via optimally windowed chirps},\ }\href@noop {} {\bibfield  {journal}
  {\bibinfo  {journal} {Phys. Rev. X}\ }\textbf {\bibinfo {volume} {8}},\
  \bibinfo {pages} {041042} (\bibinfo {year} {2018})}\BibitemShut {NoStop}%
\bibitem [{\citenamefont {Bouzid}\ \emph {et~al.}(2018)\citenamefont {Bouzid},
  \citenamefont {Keshavarz}, \citenamefont {Geri}, \citenamefont {Divoux},
  \citenamefont {Del~Gado},\ and\ \citenamefont
  {McKinley}}]{Bouzid2018JofRheology}%
  \BibitemOpen
  \bibfield  {author} {\bibinfo {author} {\bibfnamefont {M.}~\bibnamefont
  {Bouzid}}, \bibinfo {author} {\bibfnamefont {B.}~\bibnamefont {Keshavarz}},
  \bibinfo {author} {\bibfnamefont {M.}~\bibnamefont {Geri}}, \bibinfo {author}
  {\bibfnamefont {T.}~\bibnamefont {Divoux}}, \bibinfo {author} {\bibfnamefont
  {E.}~\bibnamefont {Del~Gado}},\ and\ \bibinfo {author} {\bibfnamefont
  {G.~H.}\ \bibnamefont {McKinley}},\ }\bibfield  {title} {\bibinfo {title}
  {{Computing the linear viscoelastic properties of soft gels using an
  optimally windowed chirp protocol}},\ }\href@noop {} {\bibfield  {journal}
  {\bibinfo  {journal} {Journal of Rheology}\ }\textbf {\bibinfo {volume}
  {62}},\ \bibinfo {pages} {1037} (\bibinfo {year} {2018})}\BibitemShut
  {NoStop}%
\bibitem [{\citenamefont {Lou}\ and\ \citenamefont
  {Mooney}(2022)}]{Lou2022NatRevChem}%
  \BibitemOpen
  \bibfield  {author} {\bibinfo {author} {\bibfnamefont {J.}~\bibnamefont
  {Lou}}\ and\ \bibinfo {author} {\bibfnamefont {D.~J.}\ \bibnamefont
  {Mooney}},\ }\bibfield  {title} {\bibinfo {title} {Chemical strategies to
  engineer hydrogels for cell culture},\ }\href@noop {} {\bibfield  {journal}
  {\bibinfo  {journal} {Nature Reviews Chemistry}\ }\textbf {\bibinfo {volume}
  {6}},\ \bibinfo {pages} {726} (\bibinfo {year} {2022})}\BibitemShut {NoStop}%
\bibitem [{\citenamefont {Chimene}\ \emph {et~al.}(2020)\citenamefont
  {Chimene}, \citenamefont {Kaunas},\ and\ \citenamefont
  {Gaharwar}}]{Chimene2020AdvMat}%
  \BibitemOpen
  \bibfield  {author} {\bibinfo {author} {\bibfnamefont {D.}~\bibnamefont
  {Chimene}}, \bibinfo {author} {\bibfnamefont {R.}~\bibnamefont {Kaunas}},\
  and\ \bibinfo {author} {\bibfnamefont {A.~K.}\ \bibnamefont {Gaharwar}},\
  }\bibfield  {title} {\bibinfo {title} {Hydrogel bioink reinforcement for
  additive manufacturing: A focused review of emerging strategies},\
  }\href@noop {} {\bibfield  {journal} {\bibinfo  {journal} {Advanced
  Materials}\ }\textbf {\bibinfo {volume} {32}},\ \bibinfo {pages} {1902026}
  (\bibinfo {year} {2020})}\BibitemShut {NoStop}%
\bibitem [{\citenamefont {Thompson}\ \emph {et~al.}(2022)\citenamefont
  {Thompson}, \citenamefont {Aktulga}, \citenamefont {Berger}, \citenamefont
  {Bolintineanu}, \citenamefont {Brown}, \citenamefont {Crozier}, \citenamefont
  {{in 't Veld}}, \citenamefont {Kohlmeyer}, \citenamefont {Moore},
  \citenamefont {Nguyen}, \citenamefont {Shan}, \citenamefont {Stevens},
  \citenamefont {Tranchida}, \citenamefont {Trott},\ and\ \citenamefont
  {Plimpton}}]{Thompson2022CompPhysComm}%
  \BibitemOpen
  \bibfield  {author} {\bibinfo {author} {\bibfnamefont {A.~P.}\ \bibnamefont
  {Thompson}}, \bibinfo {author} {\bibfnamefont {H.~M.}\ \bibnamefont
  {Aktulga}}, \bibinfo {author} {\bibfnamefont {R.}~\bibnamefont {Berger}},
  \bibinfo {author} {\bibfnamefont {D.~S.}\ \bibnamefont {Bolintineanu}},
  \bibinfo {author} {\bibfnamefont {W.~M.}\ \bibnamefont {Brown}}, \bibinfo
  {author} {\bibfnamefont {P.~S.}\ \bibnamefont {Crozier}}, \bibinfo {author}
  {\bibfnamefont {P.~J.}\ \bibnamefont {{in 't Veld}}}, \bibinfo {author}
  {\bibfnamefont {A.}~\bibnamefont {Kohlmeyer}}, \bibinfo {author}
  {\bibfnamefont {S.~G.}\ \bibnamefont {Moore}}, \bibinfo {author}
  {\bibfnamefont {T.~D.}\ \bibnamefont {Nguyen}}, \bibinfo {author}
  {\bibfnamefont {R.}~\bibnamefont {Shan}}, \bibinfo {author} {\bibfnamefont
  {M.~J.}\ \bibnamefont {Stevens}}, \bibinfo {author} {\bibfnamefont
  {J.}~\bibnamefont {Tranchida}}, \bibinfo {author} {\bibfnamefont
  {C.}~\bibnamefont {Trott}},\ and\ \bibinfo {author} {\bibfnamefont {S.~J.}\
  \bibnamefont {Plimpton}},\ }\bibfield  {title} {\bibinfo {title} {Lammps - a
  flexible simulation tool for particle-based materials modeling at the atomic,
  meso, and continuum scales},\ }\href@noop {} {\bibfield  {journal} {\bibinfo
  {journal} {Computer Physics Communications}\ }\textbf {\bibinfo {volume}
  {271}},\ \bibinfo {pages} {108171} (\bibinfo {year} {2022})}\BibitemShut
  {NoStop}%
\bibitem [{\citenamefont {Stillinger}(1995)}]{Stillinger1995Science}%
  \BibitemOpen
  \bibfield  {author} {\bibinfo {author} {\bibfnamefont {F.~H.}\ \bibnamefont
  {Stillinger}},\ }\bibfield  {title} {\bibinfo {title} {A topographic view of
  supercooled liquids and glass formation},\ }\href@noop {} {\bibfield
  {journal} {\bibinfo  {journal} {Science}\ }\textbf {\bibinfo {volume}
  {267}},\ \bibinfo {pages} {1935} (\bibinfo {year} {1995})}\BibitemShut
  {NoStop}%
\bibitem [{\citenamefont {Lees}\ and\ \citenamefont
  {Edwards}(1972)}]{LeesEdwards}%
  \BibitemOpen
  \bibfield  {author} {\bibinfo {author} {\bibfnamefont {A.~W.}\ \bibnamefont
  {Lees}}\ and\ \bibinfo {author} {\bibfnamefont {S.~F.}\ \bibnamefont
  {Edwards}},\ }\bibfield  {title} {\bibinfo {title} {The computer study of
  transport processes under extreme conditions},\ }\href@noop {} {\bibfield
  {journal} {\bibinfo  {journal} {J. Phys. C: Solid State Phys.}\ }\textbf
  {\bibinfo {volume} {5}},\ \bibinfo {pages} {1921} (\bibinfo {year}
  {1972})}\BibitemShut {NoStop}%
\bibitem [{\citenamefont {Stukowski}(2010)}]{ovito}%
  \BibitemOpen
  \bibfield  {author} {\bibinfo {author} {\bibfnamefont {A.}~\bibnamefont
  {Stukowski}},\ }\bibfield  {title} {\bibinfo {title} {{Visualization and
  analysis of atomistic simulation data with OVITO-the Open Visualization
  Tool}},\ }\href@noop {} {\bibfield  {journal} {\bibinfo  {journal}
  {{Modelling and Simulation in Materials Science and Engineering}}\ }\textbf
  {\bibinfo {volume} {{18}}} (\bibinfo {year} {{2010}})}\BibitemShut {NoStop}%
\bibitem [{\citenamefont {Thompson}\ \emph {et~al.}(2009)\citenamefont
  {Thompson}, \citenamefont {Plimpton},\ and\ \citenamefont
  {Mattson}}]{Thompson2009JCP}%
  \BibitemOpen
  \bibfield  {author} {\bibinfo {author} {\bibfnamefont {A.~P.}\ \bibnamefont
  {Thompson}}, \bibinfo {author} {\bibfnamefont {S.~J.}\ \bibnamefont
  {Plimpton}},\ and\ \bibinfo {author} {\bibfnamefont {W.}~\bibnamefont
  {Mattson}},\ }\bibfield  {title} {\bibinfo {title} {{General formulation of
  pressure and stress tensor for arbitrary many-body interaction potentials
  under periodic boundary conditions}},\ }\href@noop {} {\bibfield  {journal}
  {\bibinfo  {journal} {The Journal of Chemical Physics}\ }\textbf {\bibinfo
  {volume} {131}},\ \bibinfo {pages} {154107} (\bibinfo {year}
  {2009})}\BibitemShut {NoStop}%
\bibitem [{\citenamefont {Bhattacharya}\ and\ \citenamefont
  {Gubbins}(2006)}]{Bhattacharya2006Langmuir}%
  \BibitemOpen
  \bibfield  {author} {\bibinfo {author} {\bibfnamefont {S.}~\bibnamefont
  {Bhattacharya}}\ and\ \bibinfo {author} {\bibfnamefont {K.~E.}\ \bibnamefont
  {Gubbins}},\ }\bibfield  {title} {\bibinfo {title} {Fast method for computing
  pore size distributions of model materials},\ }\href@noop {} {\bibfield
  {journal} {\bibinfo  {journal} {Langmuir}\ }\textbf {\bibinfo {volume}
  {22}},\ \bibinfo {pages} {7726} (\bibinfo {year} {2006})}\BibitemShut
  {NoStop}%
\bibitem [{\citenamefont {Cates}\ \emph {et~al.}(2004)\citenamefont {Cates},
  \citenamefont {Fuchs}, \citenamefont {Kroy}, \citenamefont {Poon},\ and\
  \citenamefont {Puertas}}]{Cates2004JPhysCondMat}%
  \BibitemOpen
  \bibfield  {author} {\bibinfo {author} {\bibfnamefont {M.~E.}\ \bibnamefont
  {Cates}}, \bibinfo {author} {\bibfnamefont {M.}~\bibnamefont {Fuchs}},
  \bibinfo {author} {\bibfnamefont {K.}~\bibnamefont {Kroy}}, \bibinfo {author}
  {\bibfnamefont {W.~C.~K.}\ \bibnamefont {Poon}},\ and\ \bibinfo {author}
  {\bibfnamefont {A.~M.}\ \bibnamefont {Puertas}},\ }\bibfield  {title}
  {\bibinfo {title} {{Theory and Simulation of Gelation, Arrest and Yielding in
  Attracting Colloids}},\ }\href@noop {} {\bibfield  {journal} {\bibinfo
  {journal} {J. Phys: Condens. Matter}\ }\textbf {\bibinfo {volume} {16}},\
  \bibinfo {pages} {S4861} (\bibinfo {year} {2004})}\BibitemShut {NoStop}%
\bibitem [{\citenamefont {Inc.}(2024)}]{Mathematica}%
  \BibitemOpen
  \bibfield  {author} {\bibinfo {author} {\bibfnamefont {W.~R.}\ \bibnamefont
  {Inc.}},\ }\href {https://www.wolfram.com/mathematica} {\bibinfo {title}
  {Mathematica, {V}ersion 14.1}} (\bibinfo {year} {2024}),\ \bibinfo {note}
  {champaign, IL}\BibitemShut {NoStop}%
\end{thebibliography}%

\newpage
\clearpage

\onecolumngrid

\begin{center}
\textbf{ \large{Supporting Information: \\ 
Inter-Species Interactions in Dual, Fibrous Gel Enable Control\\ of Gel 
Structure and Rheology}}
\end{center}

\setcounter{figure}{0}
\setcounter{equation}{0}
\setcounter{paragraph}{0}
\renewcommand{\theequation}{S\arabic{equation}}
\renewcommand{\thefigure}{S\arabic{figure}}

\section*{Methods}
The model for each network and the gel preparation are based on refs.~\cite{Colombo2013PRL,Colombo2014JofRheology,Bantawa2021JPhysCondMat}.
Here, we briefly recapitulate few key ideas. The discussion of the two species and of the inter-species interactions used in the manuscript is instead new. 

\paragraph{Energy Function}
The gels are made of colloidal particles interacting via two-body ($E_2$) and three-body interaction potentials ($E_3$)~\cite{Colombo2013PRL}, 
\begin{equation}
\begin{aligned}
E(\vec{r}_1,\cdots,\vec{r}_N) = &\sum_{i=1}^{N-1}\sum_{j=i+1}^{N} E_2(r_{ij}) + \\
&\sum_{i=1}^{N}\sum_{j\ne i}\sum_{k>j \& k\ne i} E_3(\vec{r}_{ij},\vec{r}_{ik}),
\label{Eq:U}
\end{aligned}
\end{equation}
where the distance between two particles, $i$ and $j$, is given by the vector $\vec{r}_{ij} = \vec{r}_j-\vec{r}_i$, whose length is $r_{ij} = |\vec{r}_{ij}|$.
The first summation extends over all pairs of particles, whereas the second one includes all triples made of three colloidal units, with $i$ being the ``central'' particle interacting with both $j$ and $k$.
The functional form of the two-body potential is,
\begin{equation}
E_\mathrm{2}(r_{ij}) = A \epsilon \Big[\Big(\frac{d_2}{r_{ij}}\Big)^{18} - \Big(\frac{d_2}{r_{ij}}\Big)^{16}\Big] H(r_{cut}-r).
\label{Eq:2body}
\end{equation}
This potential features a sharp repulsive core, a minimum at $(9/8)^{1/2} d_2 \approx 1.06 d_2$ of energy $E_\mathrm{2}(r_{min}) \approx -0.043 A \epsilon$, and a short attractive tail that ends at $r_{cut} = 2 d$, where the interaction potential is set to zero, as shown with the step function $\Theta(x)$ ($=1$ for $x>0$ and $=0$ otherwise).
Clearly, $A$ represents the strength of attraction, and for intra-species interactions with set $A=23$ so that $E_\mathrm{2}(r_{min}) \approx - \epsilon$.
We set $d_2 = 0.922 d$, and thus the minimum of the potential is at $\approx 0.98 d$, and hence the diameter of the colloidal particles are roughly $d$.
The three-body potential is given by,
\begin{equation}
\begin{aligned}
E_\mathrm{3}(\vec{r}_{ij},\vec{r}_{ik}) &= B \epsilon \Lambda(r_{ij})\Lambda(r_{ik}) e^{-\Big[\frac{\cos(\theta_{ijk})-\cos(\overline{\theta})}{w}\Big]^2}.\\
\Lambda(r) &= \Big(\frac{d_3}{r_{ij}}\Big)^{10}\Big[1-\Big(\frac{r}{r_{cut}}\Big)^{10}\Big]^2 H(r_{cut}-r)
\end{aligned}
\label{Eq:3body}
\end{equation}
Here, $B$ is the strength of the three-body repulsion, and we set it to $B=10$ for intra-species interactions.
The value $\theta_{i,j,k}$ formed by the vectors connecting beads $i$ and $j$, and $i$ and $k$. 
The parameter $\overline{\theta}$ establishes the orientation of the beads for which the potential is minimum, and $w$ reflects the allowed deviation from this ground state.
The function $\Lambda(r)$ ensures that the repulsion decays rapidly, over a length scale given by $d_3$.
Following the standard protocol, $\overline{\theta}\approx 65^o$, or $\cos(\overline{\theta})=0.4226$, $w=0.3$, and $d_3 = 1.1d$.

\paragraph{Two Species} 
All systems are made by two colloidal species and the strength of interaction between the components are governed by the parameters $A_{ij}$ and $B_{ij}$, referring to two-body interactions and three-body interactions involving particles of species $i$ and $j$.
We note here that the three-body potential could be defined as $B_{ijk}$, as it depends on the species of three particles involved.
However, for simplicity $B_{iij}=B_{iji}=B_{jii}$, and thus it sufficies to specify only two indexes. 
Intra-species interactions are identical and equal to $A_{ii} = 23$ and $B_{ii} = 10$. 
We explore the dependence of system structure and rheology on two parameters: $\chi_2$ and $\chi_3$, which identify the strength of two-body and three-body inter-species interactions relative to the corresponding intra-species parameters.
The parameters are described in the main text.

\paragraph{Simulations Details and Gel Preparation} 
We used the same protocol described in the past~\cite{Colombo2014JofRheology}.
Briefly, the colloidal particles are randomly placed in a cubic box of side $L$ and volume $V=L^3$.
In a first round of simulations we prepare the system at a high-temperature $T_h=0.5 \epsilon/k_B$, which is large enough to prevent assembly and facilitate randomization of the initial positions.
Next, we quench the kinetic energy to a temperature $T=0.5 \epsilon/k_B$ at a rate of $9\cdot 10^{-5} \epsilon/(k_B\tau)$, down to $T_l = 0.05 \epsilon/k_B$.
Self-assembly of the fibrous networks begins during this initial quenching.
After running simulations for $5\cdot 10^3 \tau$ at $T_l$, we drain the kinetic energy for at least $15\cdot 10^3 \tau$ so that the temperature is below $10^{-9}\epsilon/k_B$, indicating that the system has reached an inherent structure.
In Fig.~S3 we compare results obtained using this protocol (red circles) or by quenching the system to much lower temperatures and repeating the calculation 4 times (triangles in shades of blue).
The differences appear negligible.
In the paper, we have either used results from the red trajectories or averages from the blue ones.

Randomization and quenching are performed using a Nos\'{e}-Hoover thermostat to control the temperature, whereas for the final quenching we adopted the following overdamped equation:
\begin{equation}
m \frac{d^2 r_{i,\alpha}}{dt^2} = F_{i,\alpha}(\vec{r}_1,\cdots,\vec{r}_N) - \zeta v_{i,\alpha},
\label{Eq:Ovdmp}
\end{equation}
where we used the $\alpha$ component of the $i$-the particle position ($r$), velocity ($v$), and force ($F$), and where the friction coefficient is $\zeta = \zeta^* \epsilon \tau/d^2$, with $\zeta^* = 1$.
All simulations are performed in version of LAMMPS~\cite{Thompson2022CompPhysComm} suitably modified to include the energy function described in the previous section. 
The integration time-step is $5\cdot 10^{-3} \tau$.

\paragraph{Modification of the Energy Function}
We tested the architecture robustness to changes in the energy function.
In order to do so, we started from a structure obtained using the above protocol, we modified the energy function and ran a new round of the overdamped dynamics described in Eq.~\ref{Eq:Ovdmp}.
We verified in all cases that again the temperature reached values around or below $10^{-9}\epsilon/k_B$.

\paragraph{Percolation}
To test if the network has percolated, we computed the clusters made by colloidal units with distance criterion $r\le 1.5d$.
Next, we triple the system size in one direction (namely, $x$), and if a cluster triples its size we call the structure percolated.
The test was performed using the clustering facility in Ovito~\cite{ovito}.

\paragraph{Start-Up Shear}
To investigate the response of the network to strain we used the start-up shear protocol 
previously described~\cite{Colombo2014JofRheology}.
Briefly, using Lees-Edwards boundary conditions~\cite{LeesEdwards}, we impose a sudden strain step of $\Delta \gamma = 0.01$ by deforming the periodic simulation box.
Next, we let the system relax for $\Delta t \approx 50 \tau$ using the zero-temperature dynamics in Eq.~\ref{Eq:Ovdmp}, and we impose another strain step.
We repeat the process until we reach a target strain.
We focus here on $\gamma \le 1$ although in some simulations larger strains were considered in order to establish the yield strain and stress.

\paragraph{Frequency Spectrum}
We used the optimally windowed chirp (OWCh) protocol~\cite{Geri2018PRX} to extract the frequency-dependent elastic ($G'$) and loss ($G''$) moduli.
The detailed implementation is described by Bouzid {\it et al}~\cite{Bouzid2018JofRheology}.
Briefly, using Lees-Edwards boundary conditions~\cite{LeesEdwards} we impose an oscillatory strain protocol, $\gamma(t)$, and we monitor the stress tensor defined by $\sigma_{\alpha\beta}(t) = -V^{-1} \sum_{i=1}^N r_{i,\alpha}(t)F_{i,\beta}(t)$, (see the LAMMPS manual and Thompson {\it et al}~\cite{Thompson2009JCP} for details).
The elastic and loss moduli are then obtained as the real and imaginary parts of $\tilde{\sigma}(\omega)/\tilde{\gamma}(\omega)$, where the symbol $\tilde{\cdots}$ refers to the Fourier transform.
During the oscillations, the dynamics of the colloidal particles is given by Eq.~\ref{Eq:Ovdmp} using the velocity relative to the movement of the solvent, which is $\dot{\gamma}(t)y_i \hat{e}_x$, where $\dot{\gamma}(t)$ is the time derivative of the strain, $y_i$ is the y-coordinate of the i-th particle, and $\hat{e}_x$ is a unit vector in the $x$ direction.

Data showing all the points up to $\omega = 0.1/\tau$ are shown in Fig.~S2.
Large fluctuations show curves that occasionally go below zero due to statistical noise.

\paragraph{Porosity} 
To compute the size of the pores in the network, we developed an algorithm inspired by Bhattacharya and Gubbins~\cite{Bhattacharya2006Langmuir} -- we follow the steps that the authors elucidated in the Introduction.
Briefly, we first create a grid of $N_g^3$ points in the simulation box, and for each one of them we compute the largest sphere, $S$, center in the grid point and such that it does not touch any of the colloidal particles of the gel.
Next, we draw $N_r$ random locations $\vec{\xi}$ in the box, we search for all the spheres $S$ that include $\vec{\xi}$, and we assign $\vec{\xi}$ to the largest of these spheres.
If the point $\vec{\xi}$ cannot be assigned because it clashes with a colloidal particle or because it is not included in any of the spheres, we reject it and draw another one.
As a result of this calculation, we construct the probability that a random position in the box unoccupied by the gel belongs to a pore larger than $R$, or $P(r>R)$.
The average pore size is $\langle R \rangle = \int_0^\infty P(r>R) dR$.
Note that the average is weighed by the volume of the pore.

\paragraph{Mixing order parameter} To identify whether the two gels have demixed, we use a demixing order parameter, $\Psi$, inspired by Cates {\it et al}~\cite{Cates2004JPhysCondMat}.
We divide the box in $N_g^3$ ($N_g=10$) boxes, and for each box $m$, we compute the number $N_m$ of particles in that box.
The order parameter is defined as,
\begin{equation}
\Psi(\{N_m\}) = \frac{\sum_{m=1}^{N_g^3} (N_m - \frac{N}{N_g^3})^2}{N \frac{N}{N_g^3}},
\label{Eq:Psi}
\end{equation}
$\Psi$ is at its minimum ($ = 0$) if all the cells are equally populated ($N_m = N/N_g^3$ for all $m$).
We have included a normalization at the denominator is chosen to be such that if half of the boxes are equally populated ($N_m = 2 N/N_g^3$) and the remaining half is empty ($N_m=0$) then $\Psi = 1$.
Note that this is not the maximum: for instance if only one third of the cells are populated by the same amount of material and the remaining two thirds are empty, then $\Phi = 2$.
Similarly, if only one cell is occupied and the rest is empty, then $\Psi = N_g^3-1$.

\paragraph{Yield strain/stress Fitting and Toughness} To find the yield strain and yield stress of the as a function of $\chi_2$, we find the maximum values of the stress ($\sigma_\mathrm{yield}$) of load curve and its corresponding strain value, $\gamma_\mathrm{yield}$. We use xmgrace to fit the the load curve to a polynomial, that is,
\begin{equation}
\sigma = a_0+ a_1 \gamma+ a_2 \gamma^2 + ......+a_n \gamma^n 
\end{equation}
where $a_n$ are fitting parameters, and ranges $n$ ranges from $5$ to $7$.
We find the maximum of that function numerically using Mathematica~\cite{Mathematica}.
Because the yield stress and strain for demixed networks were noisy, we repeated the calculation by generating four independent structures and loading curves.
For each curve, we extract the average yield stress and strain as well as the corresponding standard error on the mean. \\
The toughness was computed by integrating the load curves up until the yielding point ($\gamma_{yield}$, $\sigma_{yield}$ ) using xmgrace. We computed the toughness for each of the 4 configurations generated. The final value is generated by evaluating the mean as well as its standard error.

\begin{figure}
\centering
\includegraphics[trim=0cm 11cm 0cm 0cm,width=\textwidth]{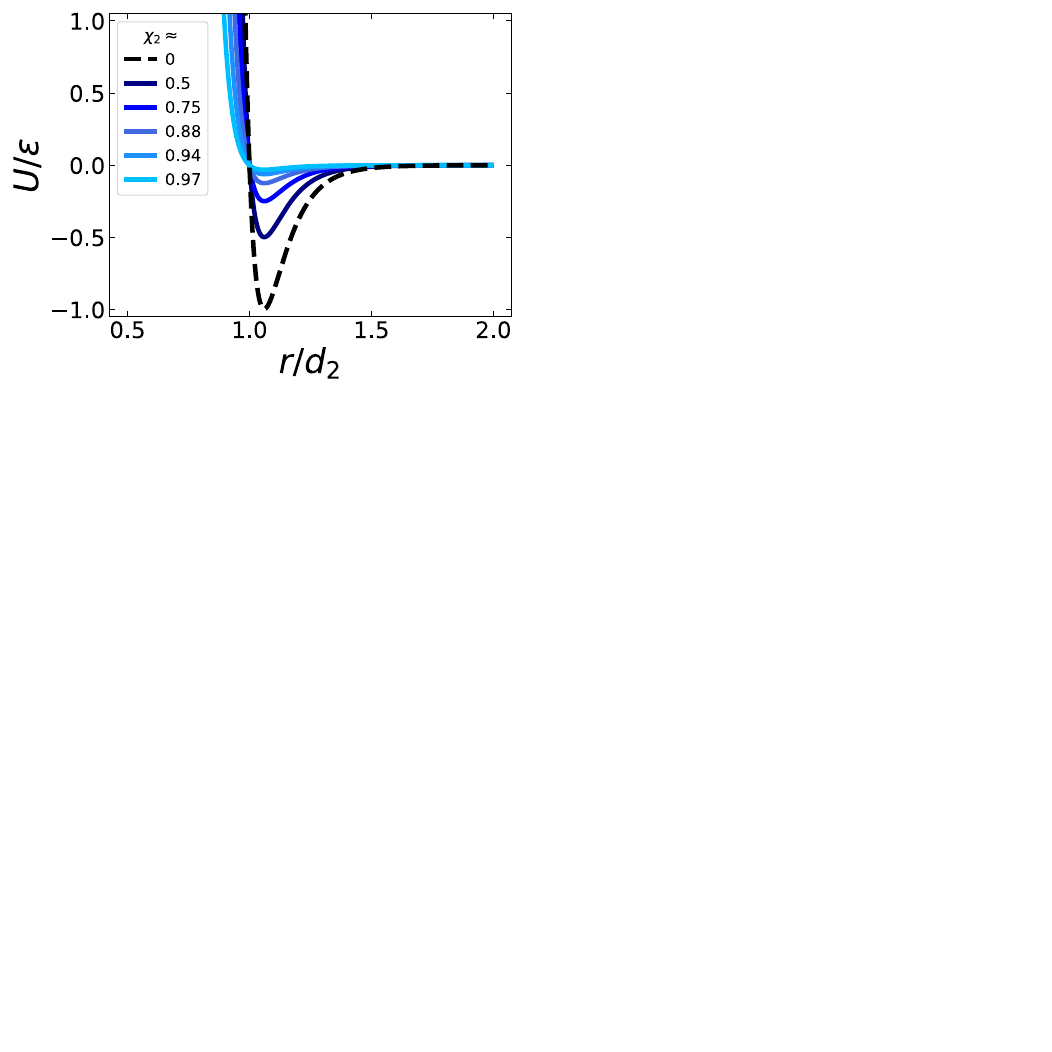}
\caption{Shape of the two-body potential as a function of $\chi_2$.
The black, dashed lines indicate the intra-species pair potential.}
\end{figure}

\begin{figure}
\centering
\includegraphics[trim=0cm 3cm 0cm 0cm,width=\textwidth]{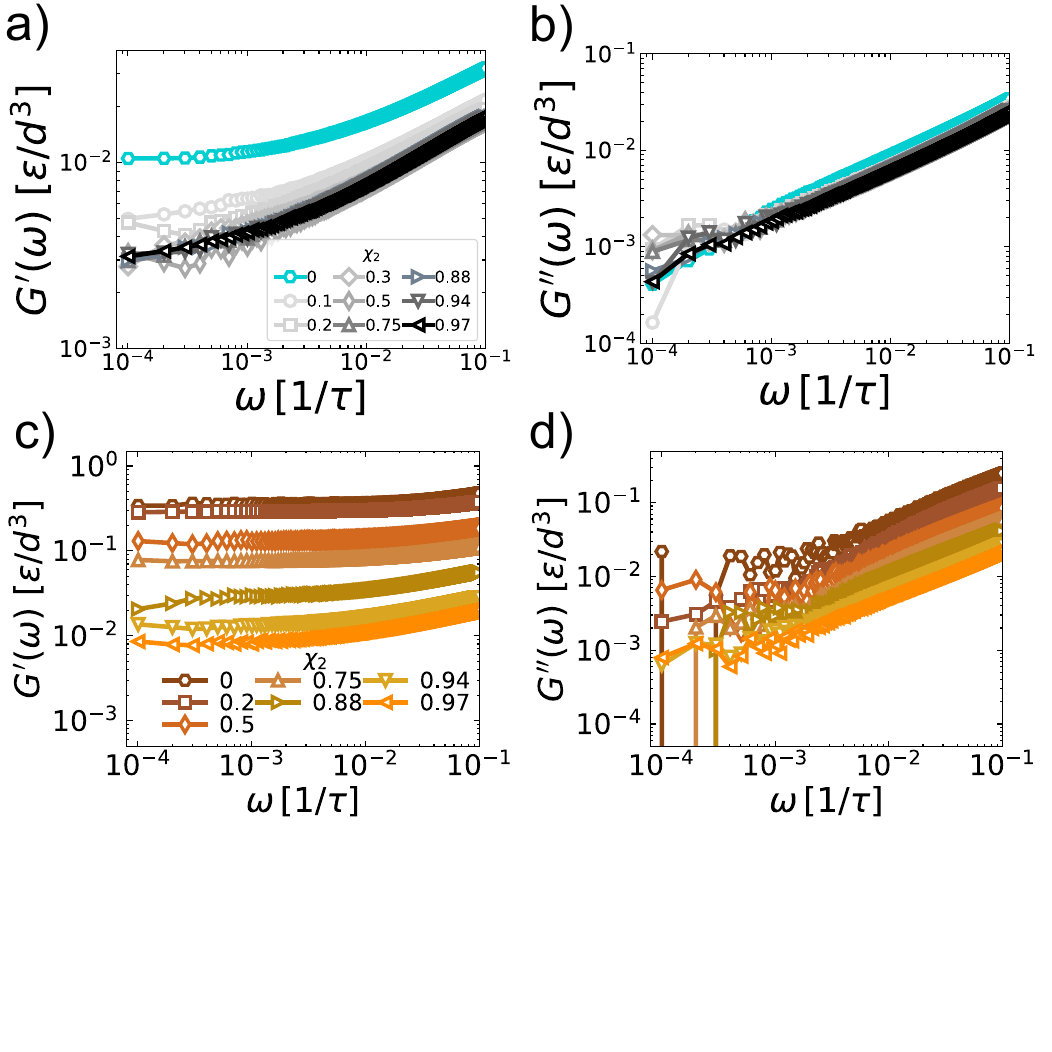}
\caption{Oscillatory linear response.
(a-b) Elastic (a) and loss (b) moduli for demixed networks as a function of $\chi_2$ (see legend where the values are approximate, the exact ones are $0$, $0.1$, $0.2$, $0.3$, $0.5$, $0.75$, $0.875$, $0.9375$, and $0.96875$).
(c-d) Same as (a) and (b), but for intertwined networks.}
\end{figure}

\begin{figure}
\centering
\includegraphics[trim=0cm 5cm 0cm 0cm,width=\textwidth]{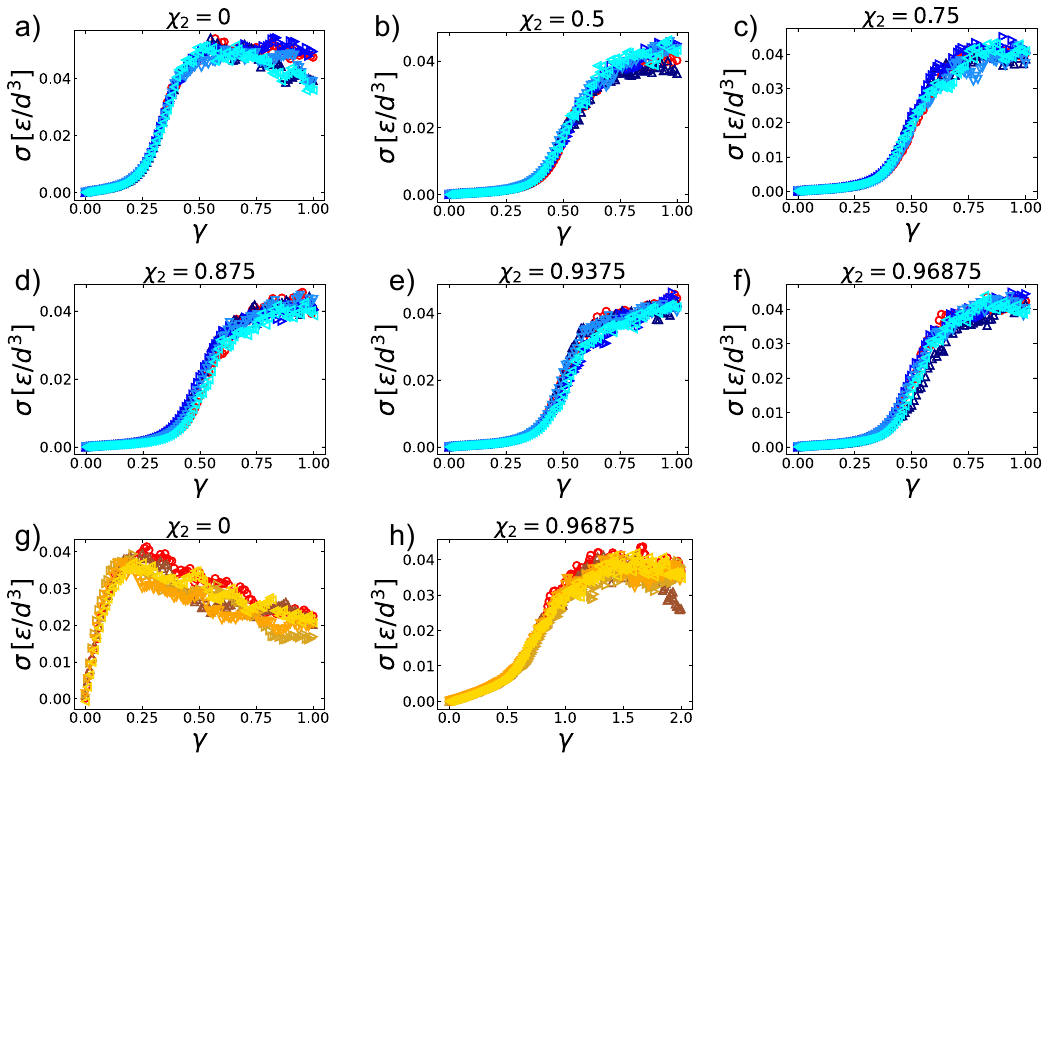}
\caption{Non-linear rheology, comparison between multiple structures.
Panels (a-f) refer to load curves for demixed networks using $\chi_2$ as shown above the figure.
The triangles in shades of blue represent four repetitions in which the temperature of the network during preparation has been quenched to extremely low values ($k_BT/\epsilon < 10^{-17}$, in most cases $<10^{-20}$).
These data were used to extract the toughness and yielding stress and strain in order to build some statistics.
The red circles represent networks that have been quenched slightly less ($k_BT/\epsilon < 10^{-9}$).
Similar results for intertwined networks are shown in panels (g-h), where the repetitions at slower quenching rate are shown in shades of orange, and the network obtained after faster quenching is in red.
The results are essentially indistinguishable.}
\end{figure}

\end{document}